\documentclass[aps,prd,10pt,amsmath,amssymb,nofootinbib,notitlepage,superscriptaddress,twocolumn,floatfix]{revtex4-2}

\usepackage[utf8]{inputenc}
\usepackage{amsmath,amssymb}
\usepackage{comment}
\usepackage{braket}
\usepackage{mathrsfs}
\usepackage{mathtools}
\usepackage{fancyhdr}
\usepackage{xcolor}
\usepackage{bm}
\usepackage{url}
\usepackage{graphicx}
\usepackage{booktabs}
\usepackage{multirow}
\usepackage{makecell}

\usepackage{ragged2e}

\usepackage{diagbox}
\usepackage{tikz}
\usetikzlibrary{patterns}
\usetikzlibrary{arrows.meta}

\allowdisplaybreaks

\definecolor{RedWine}{rgb}{0.743,0,0}
\definecolor{RoyalBlue}{rgb}{0.25,.41,.88}
\definecolor{ForestGreen}{rgb}{.13,.54,.13}

\usepackage{tensor}
\usepackage{slashed}
\usepackage{dsfont} 
\usepackage{bbold}

\makeatletter
\def\@bibdataout@aps{%
 \immediate\write\@bibdataout{%
  @CONTROL{%
   apsrev42Control%
   \longbibliography@sw{%
    ,author="48",editor="1",pages="0",title="0",year="1"%
   }{%
    ,author="48",editor="1",pages="0",title="",year="1"%
   }%
  }%
 }%
 \if@filesw
  \immediate\write\@auxout{\string\citation{apsrev42Control}}%
 \fi
}%
\makeatother

\usepackage[breaklinks,colorlinks,citecolor=ForestGreen]{hyperref}

\newcommand{\dd}{d}

\newcommand{\tauo}{\tau_0}

\newcommand{\NFE}{\mathrm{NFE}}

\newcommand{\EM}{\mathrm{EM}}
\newcommand{\Birth}{\mathscr{B}}
\newcommand{\birth}{\mathrm b}
\newcommand{\calE}{\mathcal{E}}
\newcommand{\calB}{\mathcal{B}}
\newcommand{\phim}{\varphi_{\mathrm{m}}}
\newcommand{\phiem}{\varphi_{\EM}}

\begin{document}

\title{Continuation of Force-Free Electrodynamics upon the loss of magnetic dominance}

\author{Morifumi Mizuno}
\affiliation{
Department of Physics, University of Arizona, Tucson, Arizona 85721, USA
}

\begin{abstract}
    Force-Free Electrodynamics (FFE) describes the evolution of the electromagnetic field in magnetically dominated plasmas, but ceases to be hyperbolic once the magnetic dominance condition $F^{ab}F_{ab}>0$ is lost. 
    We demonstrate that, after the loss of magnetic dominance, FFE may be replaced by a theory of null fields, $F^{ab}F_{ab}=0=F^{ab}\tilde{F}_{ab}$, characterized by the condition that its principal null direction is tangent to a geodesic congruence. 
    In flat spacetime, this theory may be equivalently stated as the condition that the field satisfies $\vec{B}^2-\vec{E}^2=0=\vec{E}\cdot\vec{B}$ and the integral curves of the drift velocity $\vec{E}\times\vec{B}/\vec{B}^{2}$ are straight lines. 
    We develop the general structure and the properties of this theory and test it against 1D PIC simulations using the collision of planar symmetric Alfv\'en waves. 
    We find that the force-free combined with the null continuation shows remarkable macroscopic agreement with PIC simulations, including the birth and evolution of the null region ($F^{ab}F_{ab}=0$) and the formation of a current sheet.
    As an independent test, we apply the null continuation to Adhikari's type-changing solution, an exact FFE solution exhibiting finite-time loss of magnetic dominance, and also find macroscopic agreement with 1D PIC simulations.  
\end{abstract}

\maketitle

\section{Introduction}
Force-free electrodynamics (FFE) \cite{Uchida1997aug,Uchida1997oct,Uchida1998jun,Komissarov2002nov,Gralla.Jacobson2014} is a universal framework derived from a simple conservation law \cite{Gralla:2018kif}, and is commonly applied to pulsar \cite{Gold1968may,GoldreichJulian1969aug} and black hole magnetospheres \cite{BlandfordZnajek1977may}. 
For the electromagnetic tensor $F_{ab}$ and current density $j^{a}$, the defining condition for FFE is the vanishing of the Lorentz force density,
\begin{align}
    \label{eq: force free condition intro}
    F_{ab}j^{b}=0.
\end{align}
The absence of the force density is what earns the theory its name \textit{force-free}. The force-free condition is a consequence of stress-energy conservation with the core assumption that the plasma is energetically negligible compared to the EM field. Under this assumption, the conservation law is imposed only on the EM stress tensor and Eq.~\eqref{eq: force free condition intro} follows naturally.  Fully deterministic FFE systems further require additional algebraic conditions, leading to the following set of equations:
\begin{align}\begin{gathered}
   F_{ab}\nabla_{c}F^{bc} = 0, \qquad \nabla_{a}\tilde{F}^{ab}=0,
   \\
    F_{ab}\tilde{F}^{ab} = 0, \qquad  F_{ab} F^{ab} > 0. \end{gathered}\label{FFE}
\end{align}
Here, $\tilde{F}_{ab}=\frac{1}{2}\epsilon_{abcd}F^{cd}$ is the dual of $F_{ab}$, and $\nabla_{a}$ is a spacetime covariant derivative.  
The first condition in Eq.~\eqref{FFE} is none other than the force-free condition \eqref{eq: force free condition intro} since $j^{a}=\nabla_{b}F^{ab}$, and the second equation is magnetic flux conservation, which also means that the two-form $F$ is closed (i.e. $dF=0$). The third one is the degeneracy condition ($\vec{E}\cdot\vec{B}=0$) ensuring the complete screening of the electric field by the plasma, and the final condition is magnetic dominance, which guarantees the evolution to be hyperbolic \cite{Komissarov2002nov,Palenzuela:2011es,Pfeiffer:2013wza,Carrasco_2016}. 

Equations~\eqref{FFE} illustrate a remarkable feature of FFE: plasma is indeed present through the current but its independent degrees of freedom (e.g. density, velocity) are absent.
Plasma is indispensable for FFE since it must supply enough charged particles to fully screen the parallel electric field, and to support current and charge for the FFE field, while remaining energetically insignificant. However, beyond that, it plays no further role in the theory. 
Therefore, FFE is a theory whose behavior is insensitive to the details of plasma composition or underlying microscopic plasma physics. 
This property of FFE is perhaps most iconically stated as \textit{plasma physics without plasma}.

Despite its elegance, FFE is not a complete theory; it contains within itself the possibility of failing to be hyperbolic by spontaneously losing magnetic dominance. 
In other words, the loss of magnetic dominance signals the loss of predictability of FFE.  
The failure of FFE is also supported by considering the plasma motion. Upon the loss of magnetic dominance, the field becomes null $F^{ab}F_{ab}=0$ or electrically dominated $F^{ab}F_{ab}<0$. However, in this case, particles can no longer remain energetically negligible. They are forced to gain energy \cite{Levinson:2022jzs,Komissarov2002nov}, violating one of the assumptions of FFE. 
Such finite-time spontaneous loss of magnetic dominance has been observed numerically \cite{Komissarov2002nov,Spitkovsky_2006} and handled in various ways (clipping of the electric field, dissipation, and resistivity in electrically dominated regions) \cite{Komissarov:2005xc,Spitkovsky_2006,McKinney:2006sc,Kalapotharakos:2008zc,Kalapotharakos2012apr,Li_2012,Parfrey_2012,Alic:2012df}. More recently, exact FFE solutions exhibiting a smooth transition to electric dominance have been derived \cite{Adhikari:2025rya,sanchez2026semilinearwavesectorforcefree}, indicating that the transition to a null or electrically dominated regime is a mathematically generic property of FFE. 
Whether the loss of predictability of FFE occurs depends on the initial data, but the key observation is that FFE has a built-in mechanism to evolve toward its own demise. 

The subsequent evolution after the loss of magnetic dominance lies outside the description of FFE. One may then abandon the idea of an effective description and return to plasma-resolved models such as magnetohydrodynamics (MHD) and particle-in-cell (PIC) simulations \cite{Paschalidis_2013,Philippov_2015,Cerutti_2015,Ripperda_2021,N_ttil__2022}. 
However, the loss of magnetic dominance does not necessarily imply that the entire program of effective theory has failed. Instead, it may merely signal the end of FFE, and the further evolution may still be described by a new effective theory without restoring plasma degrees of freedom. This possibility of going beyond FFE while keeping the theory minimal has been considered in several forms \cite{Gruzinov:2013pva,Jacobson:2015cia,Gralla:2018kif,Gralla:2018bvg}. In this paper, we take this perspective as our guiding principle and we seek a new plasma-free theory that extends beyond FFE. 
Here, we argue that such a successor theory \textit{does exist}, and it takes over from FFE precisely when $F^{ab}F_{ab}=0$ is reached.

Our proposed theory is \textit{Null-Force Electrodynamics (NFE)}. It is a theory describing a null electromagnetic field ($\tilde{F}_{ab}F^{ab}=0=F^{ab}F_{ab}$), and its appearance is triggered by the loss of magnetic dominance in FFE. Unlike FFE, the Lorentz force density for NFE is no longer required to vanish; instead, the defining condition for NFE is the \textit{null-force} condition:
\begin{align}
    \label{eq: null force condition intro}
    F\indices{^a_b}j^{b}\propto \ell^{a}.
\end{align}
Here, $\ell$ is the unique (up to scaling) principal null direction (PND) of the null field  \cite{Tafel1985jul,Gralla.Jacobson2014,Menon2021} satisfying $F_{ab}\ell^{b}=0$. The presence of a nonzero Lorentz force density implies that NFE allows the exchange of EM energy-momentum with matter. But Eq.~\eqref{eq: null force condition intro} forces such interaction to occur only along the null direction $\ell$. This is the sense in which the theory is ``null-force''. 

The 3+1 form of the null-force condition \eqref{eq: null force condition intro} is also informative. 
In 3+1D flat spacetime, the components of the PND of the null field ($\vec{E}\cdot\vec{B}=0=\vec{B}^2-\vec{E}^2$) may be written as $\ell^{\mu}=(1,\vec{\ell})$ where the spatial part is $\vec{\ell}=(\vec{E}\times\vec{B})/\vec{B}^2$. If we write the proportionality constant in Eq.~\eqref{eq: null force condition intro} as $C$, the time component of Eq.~\eqref{eq: null force condition intro} fixes it to be $C=\vec{j}\cdot\vec{E}$, provided that $\vec{j}$ and $\rho$ are the current and charge densities. Then, the spatial component of Eq.~\eqref{eq: null force condition intro} may be given by
\begin{align}
    \label{eq: 3+1 null force field only intro}
    \rho \vec{E}+\vec{j}\times\vec{B}&=(\vec{j}\cdot\vec{E})\frac{\vec E\times \vec B}{\vec B^{\,2}}.
\end{align}
In the 3+1 language, $\vec{\ell}=(\vec{E}\times\vec{B})/\vec{B}^2$ is a ``drift velocity'' (for the null field, the drift velocity reaches the speed of light, i.e. $\vec{\ell}\cdot\vec{\ell}=1$).  Therefore, the condition \eqref{eq: 3+1 null force field only intro} may be stated as the Lorentz force density pointing in the direction of the drift velocity with its magnitude $\vec{j}\cdot\vec{E}$. 
Unlike FFE, generic NFE fields have $\vec{j}\cdot\vec{E}\neq0$, manifesting the fact that the field exchanges energy with the plasma.

A formulation of NFE parallel to the field-only formulation of FFE \eqref{FFE} may also be obtained. By contracting Eq.~\eqref{eq: null force condition intro} with $F_{ca}$, the right-hand side vanishes, resulting in $F_{ca}F\indices{^a_b}j^{b}=0$. As we discuss in Sec.~\ref{sec:nfe}, Eq.~\eqref{eq: null force condition intro} can be recovered from this condition, provided that the field is null and closed. Therefore, an equivalent way of defining NFE is the set of equations:
\begin{align}\begin{gathered}
      F_{ca}F\indices{^a_b}\nabla_{d}F^{bd} = 0, \qquad  
    \nabla_{a}\tilde{F}^{ab}=0,
      \\ F_{ab}\tilde{F}^{ab} = 0, \qquad  F_{ab} F^{ab} = 0. \end{gathered}\label{NFE}
\end{align}

Equations \eqref{eq: 3+1 null force field only intro} and \eqref{NFE} show a clear similarity between FFE and NFE: no independent plasma degrees of freedom appear. Plasma is essential but its microscopic details are irrelevant. Therefore, the iconic property of FFE -- \textit{plasma physics without plasma} -- also exists in NFE. However, NFE is not a generalization of FFE; it is a separate theory defined on the null regime $F^{ab}F_{ab}=0$. In this sense, FFE and NFE have equal status, both providing plasma-free effective descriptions applicable to different electromagnetic regimes.

The main purpose of this paper is to develop NFE, explore its various properties, and test the extent of the validity of this description. We do so in a fully covariant language. As we will show,  NFE has various geometric properties, and the spacetime perspective enables us to identify its intrinsic nature and avoid introducing arbitrary-frame-dependent objects. 

The first notable finding about NFE is that, like FFE, NFE follows from a simple conservation law. 
The difference from FFE is that the conservation law needs to be imposed for the \textit{entire} stress-energy tensor, rather than the EM stress tensor alone. This is a natural consequence of the fact that, once the field becomes null, the matter stress tensor is not negligible. 
The key assumption leading to NFE is that the matter stress tensor takes the form of null dust flowing along the direction of the PND. 
This applies to a wide range of plasmas since, in the null field limit, the leading-order behavior of charged particles is to flow in the direction of $\ell$ regardless of microscopic details. In NFE, the energetic contribution of matter is not an independent variable; instead, it needs to be determined to support the NFE field.
Thus, once NFE is solved, the macroscopic matter stress-energy tensor may be obtained \textit{solely} in terms of the EM field. 

NFE shares several properties present for null FFE \cite{Menon2021,Menon:2026xnj}, including the existence of field sheets (two-dimensional manifolds whose tangent spaces are the kernel of $F$ \cite{Gralla.Jacobson2014}). In null FFE, field sheets are null and the PND is tangent to a null geodesic congruence \cite{Menon:2026xnj,Menon2021,Brennan:2013kea}. We show that this is also true for NFE. A special property intrinsic to NFE is that a local converse is also true: for a given local null geodesic congruence, we can construct a wide class of NFE solutions whose PND agrees with the tangent to the chosen congruence. 
Furthermore, we clarify the geometric relation between NFE and null FFE in terms of the shear of the congruence. For null FFE, the congruence and field sheet foliation need to be chosen such that a shear component aligned with the field sheet vanishes \cite{Menon:2026xnj}, while for NFE no such restriction is needed.

The same geometric viewpoint gives a simple method of finding an NFE solution. We first provide an initial value formulation of NFE, namely a specification of the field data on a spacelike hypersurface $\Sigma$. Then, we find that using the ``method of characteristics'' (e.g. \cite{evans10}) enables us to find the NFE solution with the specified initial data by Lie transporting the data along the null geodesic congruence launched from the initial surface. This reflects the fact that the NFE field is governed by Lie transport along the null congruence.

A generic null congruence can eventually form a conjugate point in a finite time. Since the field is transported along the congruence, the formation of conjugate points signals the breakdown of the smooth NFE description and the formation of a \textit{current sheet}. This is another notable property of NFE: it has predictive power for the onset of singularities. 
The subsequent evolution after the current sheet formation cannot be determined solely by NFE since a physical current sheet is a thin kinetic layer whose dynamics require the details of microscopic plasma physics \cite{Sironi_2014,Cerutti_2015,Sironi:2025kgn}. 
In this paper, we do not model the internal structure of the current sheet. Instead, following \cite{Li_Xinyu_Beloborodov_2021}, we treat the current sheet as a boundary condition motivated by the plasma screening of the electric field inside the sheet and formulate its condition covariantly.

For astrophysically relevant evolution of FFE, the loss of magnetic dominance occurs in some locations in spacetime (e.g. current sheets in pulsars and collisions of Alfv\'en waves \cite{Li_Xinyu_Beloborodov_2021,N_ttil__2022,Ripperda_2021}) while other regions remain in FFE. Thus, FFE and NFE generally coexist, separated by moving FFE/NFE interfaces. We show that, as with the current sheet, the physical interface is a kinetic layer, which in general carries a nonzero surface stress tensor.

To demonstrate how NFE appears in practice, we study two analytical examples of FFE solutions that exhibit temporal type-changing behavior. 
The first one is the collision of planar symmetric, counter-polarized (i.e. the transverse magnetic fields of the incoming waves point in opposite directions) Alfv\'en waves. We choose this example since it may have direct applications to astrophysics \cite{Li_Xinyu_Beloborodov_2021,N_ttil__2022,Ripperda_2021}. 
We take the incoming waves to have a triangular form and analyze the collision in \(1+1\) dimensions.
We identify the NFE region and find the analytical expression of the field therein. 
Furthermore, the full collision process, including the current sheet formation and the evolution of the FFE/NFE interface, is computed by using the current sheet boundary condition and by specifying a physically reasonable interface closure. 
The constructed field in the FFE+NFE description is compared with a 1+1D PIC simulation of the same Alfv\'en wave collision using Tristan-v2 \cite{tristan_v2}. We show that the FFE+NFE model exhibits a close agreement with the PIC simulations on a macroscopic scale.

The second example of an FFE solution is Adhikari's type-changing solution \cite{Adhikari:2025rya}. Our choice of this example is motivated by the fact that it provides a much cleaner picture of FFE+NFE theory. 
In this FFE model, the transition to an electrically dominated region occurs smoothly on the whole Cauchy surface. In this case, the whole FFE region is replaced by the NFE region and no FFE/NFE interface is formed; thus, no modeling for the interface law is required. 
Furthermore, we show that the resulting NFE continuation does not form a current sheet, so no current sheet modeling is needed. Thus, the entire evolution of the system can be found fully within the FFE+NFE description with no prescription for the current sheet or the FFE/NFE interface. 
We also compare the FFE+NFE extension against a 1+1D PIC simulation and show that the FFE+NFE extension agrees with PIC on a macroscopic scale.

This paper is organized as follows.  In Sec.~\ref{sec:ffe-review}, we quickly review FFE.  In Sec.~\ref{sec:nfe}, we define NFE, show how it follows from total stress-energy conservation, explain its connection to null geodesic congruences, and present a method to find solutions from specified initial data as well as Ohm's law for NFE in 3+1 form. In Sec.~\ref{sec:current sheet interface}, we discuss current sheet formation and the FFE/NFE interface condition.  In Sec.~\ref{sec:applications}, we apply the framework to two examples: the collision of planar symmetric Alfv\'en waves and Adhikari's type-changing force-free solution.  The solutions are constructed and compared with the PIC simulation results.
In Sec.~\ref{sec:limitations}, we discuss the limitations of the effective description and in Sec.~\ref{sec:outlook}, we outline possible future directions.

Our conventions are as follows. Latin indices $a,b,c\dots$ are abstract tensor indices, while Greek indices $\mu,\nu, \dots$ are tensor components that run from $0$ to $3$. Our metric has signature $(-,+,+,+)$ and $c=1$. The orientation of the field tensor is $F_{xy}=+B_{z}$.

\section{Short review of Force Free Electrodynamics}\label{sec:ffe-review}

Let $F$ be the electromagnetic field tensor, satisfying the Maxwell equations:
\begin{align}
    \label{eq: dF=0}
    \nabla_{[a}F_{bc]}&=0,
    \\
    \label{eq: nabla F=j}
    \nabla_{b}F^{ab}&=j^{a},
\end{align}
where the square brackets denote the antisymmetrization of indices (e.g. \cite{Wald:1984rg}) and $\nabla_{a}$ is the spacetime covariant derivative. 

FFE is an effective description of the evolution of EM fields in the presence of plasma. The central assumptions of FFE are (i) a degenerate field, (ii) magnetic dominance, and (iii) that the plasma energy is negligible compared to the energy of the EM field.
The degeneracy and magnetic dominance conditions are respectively written as 
\begin{align}
     \label{eq FFE degenerate}
    \tilde{F}^{ab}F_{ab}&=0,
    \\
    \label{eq: FFE magnetic dominance}
    F^{ab}F_{ab}&>0.
\end{align}
The first condition implies that no parallel electric field exists (i.e. $\vec{E}\cdot\vec{B}=0$), and the second condition ($\vec{B}^{2}-\vec{E}^{2}>0$) is required for FFE to be hyperbolic \cite{Komissarov2002nov,Pfeiffer:2013wza,Carrasco_2016}. 

A degenerate and closed (i.e. $dF=0$ or equivalently Eq.~\eqref{eq: dF=0}) two-form has the geometric property that its kernel is involutive. By the Frobenius theorem, an involutive vector space has a corresponding submanifold to which it is tangent. In our case, the submanifold represents the world sheet swept by the magnetic field line, and the magnetic dominance condition ensures that it is timelike. Such a submanifold is referred to as a \textit{field sheet} \cite{Gralla.Jacobson2014}. Since the guiding center velocity $v^{a}$ of charged particles in a magnetically dominated field satisfies $F_{ab}v^{b}=0$ \cite{Kruskal1958,Northrop1961,Vandervoort1960jul}, a particle's velocity is tangent to the field sheet, or we say \textit{the particle is stuck to the field line}. In the frame of a particle stuck to the field line, the electric field completely vanishes, and no strong acceleration occurs.

The third assumption provides the dynamical equation defining FFE. 
Let $T_{\rm EM}^{ab}$ and $T_{\rm m}^{ab}$ denote the electromagnetic and matter stress-energy tensors. Conservation of the total stress-energy tensor gives
\begin{align}
    \label{eq:total energy conservation}
    \nabla_a\left(T_{\rm EM}^{ab}+T_{\rm m}^{ab}\right)=0,
\end{align}
where the electromagnetic stress-energy tensor is
\begin{align}
    \label{eq: T_EM}
    T_{\rm EM}^{ab}=F\indices{^{a}_{c}}F^{bc}-\frac{1}{4}F^{cd}F_{cd}g^{ab}
\end{align}
In the force-free limit, the electromagnetic field is assumed to carry the dominant energy and momentum. This assumption implies that the matter stress-energy tensor can be dropped in Eq.~(\ref{eq:total energy conservation}), so the electromagnetic stress-energy tensor is conserved by itself:
\begin{align}
    \label{eq: nabla T_Em=0}
    \nabla_a T_{\rm EM}^{ab}=0.
\end{align}
Using Eq.~\eqref{eq: dF=0} and Eq.~\eqref{eq: T_EM}, Eq.~\eqref{eq: nabla T_Em=0} leads to the force-free condition,
\begin{align}
    \label{eq: force free condition}
    0&=\nabla_{b}T^{ab}_{\rm EM}=-F\indices{^{a}_{b}}j^b.
\end{align}
If the whole effect of matter is completely ignored, i.e. $j^{a}=0$, vacuum Maxwell equations are recovered. However, if $j^{a}$ remains nonzero, Eq.~\eqref{eq: force free condition} together with Eq.~\eqref{eq: dF=0} provides nonlinear equations solely in terms of $F$:  
\begin{align}
    \label{eq: define FFE}
   \nabla_{[a}F_{bc]}&=0,\quad F\indices{^{a}_{b}}\nabla_{c}F^{bc}=0.
\end{align}
These are the defining equations for FFE. This system may first be cast into an evolution form \cite{gruzinov1999stabilityforcefreeelectrodynamics}, whose solution then can be found numerically \cite{Spitkovsky_2006}, provided magnetic dominance holds.

There is a version of FFE called ``null FFE'' \cite{Menon2021} in which the magnetic dominance condition is replaced by the null condition $F^{ab}F_{ab}=0$. In this paper, our definition of FFE strictly demands magnetic dominance \eqref{eq: FFE magnetic dominance}, so null FFE is technically excluded. This is because null FFE and the usual FFE have qualitatively distinct properties; for example, for the usual FFE, the field sheet is timelike, but for null FFE, it is null. Also, null FFE is closely tied to null geodesics \cite{Menon2021}. We will discuss null FFE in the context of NFE in Sec.~\ref{subsec: null FFE is subsector of NFE}.

\section{Null-Force Electrodynamics}\label{sec:nfe}

Once \(F^{ab}F_{ab}=0\) is reached, FFE is no longer valid, and we replace the subsequent evolution of the field using NFE. We first provide the definition and basic properties of NFE. We will discuss the validity of NFE in Sec.~\ref{subsec:energy conservation} in the context of the derivation of NFE from the conservation law. 

In NFE, the electromagnetic field is null, which is defined by
\begin{align}
    \label{eq: null electromagnetic field}
    F^{ab}F_{ab}=0,\qquad F_{ab}\tilde{F}^{ab}=0.
\end{align}
For a nonzero null field, there is a unique PND
\(\ell^a\), up to rescaling, satisfying \cite{Tafel1985jul,Gralla.Jacobson2014,Menon2021}
\begin{align}
    \label{eq: Fell=0}
    F_{ab}\ell^{b}=0.
\end{align}
For a generic null field with PND $\ell$, it can be shown that (e.g.~\cite{Menon:2026xnj}) the field takes the following form: 
\begin{align}
    \label{eq: canonical null F}
    F_{ab}=2\ell_{[a}q_{b]},
\end{align}
for some vector $q^{a}$ which is orthogonal to $\ell$ ($q\cdot\ell=0$ where $\cdot$ denotes contraction). 
Using Eqs.~\eqref{eq: T_EM} and \eqref{eq: canonical null F}, the electromagnetic stress-energy tensor takes the null-dust form
\begin{align}
    \label{eq: T_em null dust}
    T_{\rm EM}^{ab}=F\indices{^{a}_{c}}F^{bc}=\varphi_{\rm EM}\, \ell^a \ell^b,
\end{align}
where \(\varphi_{\rm EM}=q^{a}q_{a}\) is a nonzero scalar characterizing the strength of the field.
In FFE, the Lorentz force density vanishes. On the other hand, in NFE, the Lorentz force density is allowed to
be nonzero, but only in the direction of the PND, which gives the defining condition for NFE:
\begin{align}
    \label{eq: null force density}
    F\indices{^{a}_{b}}j^{b}\propto \ell^{a}.
\end{align} 
The first observation about NFE is that $\ell$ is \textit{tangent to a null geodesic}. 
To see this, use the identity $\nabla_{b}T_{\rm EM}^{ab}=-F\indices{^a_b}j^{b}$ with Eq.~\eqref{eq: T_em null dust}. Then,
\begin{align}
    \ell^{a}&\propto F\indices{^a_b}j^{b}\notag
    \\
    &=-\nabla_{b}T_{\rm EM}^{ab}\notag
    \\
    &=-\nabla_{b}\left(\varphi_{\rm EM}\, \ell^a \ell^b\right)\notag
    \\
    &=-\nabla_{b}\left(\varphi_{\rm EM}\, \ell^b\right)\ell^a- \varphi_{\rm EM}\, \ell^b\nabla_{b}\ell^{a}.
\end{align}
From this, $\ell^a$ satisfies 
\begin{align}
    \ell^{b}\nabla_{b}\ell^{a}&\propto \ell^{a},
\end{align} 
meaning that $\ell^{a}$ is null geodesic tangent. 
We use the remaining scaling freedom in $\ell^a$ to choose an affine parametrization, so that
\begin{align}
    \label{eq: affine null geodesics}
    \ell^{a}\nabla_{a}\ell^{b}=0.
\end{align}
In fact, we can also show that the converse holds. For a closed null field whose PND is tangent to a null geodesic, the Lorentz force density is
\begin{align}
    \label{eq:geodesic to null force}
    F\indices{^{a}_{b}}j^{b}&=-\nabla_{b}T_{\rm EM}^{ab}\notag
    \\
    &=-\nabla_b\left(\phiem\ell^{a}\ell^{b}\right)\notag
    \\
    &=-\nabla_{b}\left(\varphi_{\rm EM}\ell^{b}\right)\ell^{a}\propto \ell^{a}.
\end{align}
Thus, the null-force condition \eqref{eq: null force density} and the condition that the PND is tangent to a null geodesic \eqref{eq: affine null geodesics} are equivalent. 

NFE can be written without explicitly introducing PND. 
Using Eq.~\eqref{eq: Fell=0}, the null-force condition \eqref{eq: null force density}
can be written as
\begin{align}
    \label{eq: NFE condition}
    F^{ab}F_{bc}j^{c} = 0.
\end{align}
In fact, this condition is equivalent to Eq.~\eqref{eq: null force density}. To see this, 
use Eq.~\eqref{eq: T_em null dust} to write $ F^{ab}F\indices{_b^c}$ as a tensor product of $\ell^{a}$ and $\ell^{c}$, then Eq.~\eqref{eq: NFE condition} implies
\begin{align}
    0=\ell^{a}(\ell\cdot j).
\end{align}
Since $\ell$ is nonzero, we find
\begin{align}
    \label{eq: j.ell=0}
    \ell \cdot j=0.
\end{align}
Using Eq.~\eqref{eq: T_em null dust} and \eqref{eq: j.ell=0}, the square of the Lorentz force density may be shown to be null since
\begin{align}
    (F\indices{^a_b}j^{b})(F_{ac}j^{c})=
    j^{b}F\indices{^a_b}F_{ac}j^{c}\propto (\ell\cdot j)^{2}=0.
\end{align}
The condition \eqref{eq: NFE condition} states that $F^{ab}j_{b}$ is tangent to the null field sheet. Since for the null field, the only null direction that annihilates the field is $\ell$, the null-force condition~\eqref{eq: null force density} follows.  
In other words, Eq.~\eqref{eq: NFE condition} together with Maxwell equations \eqref{eq: nabla F=j} and \eqref{eq: dF=0} as well as the null condition \eqref{eq: null electromagnetic field} gives the field-only form of NFE shown in \eqref{NFE} in the Introduction. The preceding discussion implies that the conditions \eqref{eq: null force density}, \eqref{eq: NFE condition}, and \eqref{eq: j.ell=0} are all equivalent, given Eqs.~\eqref{eq: dF=0} and \eqref{eq: null electromagnetic field} \footnote{
We have shown Eq.~\eqref{eq: null force density} $\Rightarrow$ Eq.~\eqref{eq: NFE condition} $\Rightarrow$ Eq.~\eqref{eq: j.ell=0} $\Rightarrow$ Eq.~\eqref{eq: null force density}, given that $F_{ab}$ is null and closed. }.

The transport structure also follows immediately. Since \(dF=0\) and
\(\ell\cdot F=0\), Cartan's formula \footnote{Cartan's formula is
\(\mathcal{L}_{v}\omega=d(v\cdot \omega)+v\cdot d\omega\), where \(v\) is a
vector, \(\omega\) is a form, and \(v\cdot\omega\) denotes contraction of
\(v^a\) into the first index of \(\omega\).} gives
\begin{align}
    \label{eq: Lie F =0}
    \mathcal{L}_\ell F=d(\ell\cdot F)+\ell\cdot dF=0.
\end{align}
Thus the electromagnetic two-form is Lie transported along null geodesics.  We will see in Sec.~\ref{sec: initial value formulation} that NFE solutions can be found simply by transporting the field along the null geodesics launched from the initial data surface. 

\subsection{NFE from conservation of total stress-energy}\label{subsec:energy conservation}

NFE can be motivated from the conservation of total
stress-energy.  The key physical assumption is that, once the FFE field approaches
null, the leading matter stress tensor is also a null dust flowing
along \(\ell^a\), namely
\begin{align}
    \label{eq: T_matter null dust}
    T_{\rm m}^{ab}=\varphi_{\rm m} \ell^{a}\ell^{b},
\end{align} 
where $\phim$ is a nonzero scalar characterizing the matter stress-tensor. At this point, we need not know the details of $\phim$. 
Then, the total stress-energy tensor in the NFE region is
\begin{align}
    \label{eq: NFE stress energy tensor}
    T^{ab}_{\rm NFE}=T^{ab}_{\rm EM}+T^{ab}_{\rm m}=\varphi_{\rm NFE}\ell^{a}\ell^{b},
\end{align}
where
\begin{align}
    \varphi_{\rm NFE}=\varphi_{\rm EM}+\varphi_{\rm m}.
\end{align}
The motivation for the assumption Eq.~\eqref{eq: T_matter null dust} is the following. In the magnetically dominated FFE region, charged matter is confined to the field sheets, so its velocity lies in the kernel of $F$. As $F^{ab}F_{ab}\to0$, the two PNDs of the magnetically dominated field degenerate into the single null direction $\ell^a$ \cite{Jacobson:2015cia}. Thus the leading form of the matter stress tensor can naturally be taken to be null dust flowing in the direction of $\ell$. The matter contribution need not vanish there, but its dominant stress-energy component has the same null-dust form as the electromagnetic stress-energy tensor.

Given the assumptions \eqref{eq: null electromagnetic field} and \eqref{eq: T_matter null dust}, the basic NFE equations follow from total stress-energy conservation. Taking the divergence of the total stress-energy \eqref{eq: NFE stress energy tensor}, we obtain
\begin{align}
    0&= \nabla_a T_{\rm NFE}^{ab}\notag
    \\
    &=\nabla_{a}\left(\varphi_{\rm NFE}\ell^{a}\ell^{b}\right)\notag
    \\
    &=\varphi_{\rm NFE}\ell^{a}\nabla_{a}\ell^{b}
    +\nabla_{a}\left(\varphi_{\rm NFE}\ell^{a}\right)\ell^{b}.
    \label{eq: NFE energy conservation}
\end{align}
In other words, we immediately obtain that $\ell^{a}$ is tangent to a null geodesic. As we did earlier, we can set $\ell$ to be affinely parametrized using the remaining scaling freedom in $\ell^a$. As we have shown in Eq.~\eqref{eq:geodesic to null force}, the null-force condition follows from the fact that the PND is tangent to a null geodesic.

The presence of a nonzero $\phim$ generally gives a nonzero Lorentz force density. To see this, notice that the right-hand side of Eq.~\eqref{eq:geodesic to null force} can be rewritten in terms of $T^{ab}_{\rm m}$ using Eq.~\eqref{eq: NFE stress energy tensor} on account of the conservation of the total stress-tensor $T^{ab}_{\rm NFE}$.  Then, since $\ell^{a}$ is affinely parametrized, we obtain
\begin{align}
    \label{eq: null force matter}
    F\indices{^a_b}j^{b}=\nabla_{b}T^{ab}_{\rm m}=\nabla_{b}(\phim \ell^{b})\ell^{a}.
\end{align}
This expression relates the matter energy and the Lorentz force, demonstrating the fact that the system is \textit{not force-free} and that NFE is a theory that is energetically involved with both the field and matter.

The derivation of NFE based on the conservation law implies a wide range of applicability of NFE. In order for NFE to be valid, we only need the field to be null \eqref{eq: null electromagnetic field} and the matter to be described by Eq.~\eqref{eq: T_matter null dust}. For the system to remain null, it has to have a sufficient amount of plasma to screen any electric fields that contribute to the strong acceleration of the plasma. The presence of enough plasma keeps the field degenerate and prevents electrically dominated regions from forming. The null dust assumption of the matter requires that the stress-tensor components other than the null flow (such as temperature or the pressure of the plasma) are energetically negligible compared to the EM field. As long as these conditions are met, we expect NFE to emerge.

\subsection{Geometry of NFE and null geodesic congruence}\label{subsec: null FFE is subsector of NFE}
NFE shares several geometric properties with null vacuum electrodynamics
\cite{Robinson1961may,Tafel1985jul} and null FFE
\cite{Menon2021,Menon:2026xnj,Brennan:2013kea}.  The key difference between NFE and null FFE is the
Lorentz force law: null FFE requires the Lorentz force density to vanish, while
NFE only requires it to be parallel to the PND $\ell$.  
To
understand the geometric properties of NFE as well as its differences from and similarities to null FFE, let us recall the geometric properties of general null fields. 
For a null
field, field sheets are null surfaces and each sheet contains
a distinguished null direction, which is identified as \(\ell^a\).  
In general, however, \(\ell^a\) need not be tangent to a null geodesic.\footnote{For example, in Minkowski spacetime,
\[
F=-\sin\vartheta(t)\,dt\wedge dx+\cos\vartheta(t)\,dt\wedge dy+dx\wedge dy
\]
is null and closed, \(F_{ab}F^{ab}=0=F_{ab}\tilde F^{ab}\) and \(dF=0\).  Its PND is
\(\ell=\partial_t-\cos\vartheta\,\partial_x-\sin\vartheta\,\partial_y\), but
\(\ell^a\nabla_a\ell=\dot\vartheta(\sin\vartheta\,\partial_x-\cos\vartheta\,\partial_y)\), which is not proportional to \(\ell\) unless \(\dot\vartheta=0\).}

We have shown in Eq.~\eqref{eq: affine null geodesics} that for any NFE field, there is a corresponding $\ell$ that is tangent to a null geodesic. We have also shown that a null field whose PND is tangent to a null geodesic is an NFE field. 
Here we establish a more general statement: for any local null geodesic congruence, one can construct NFE solutions whose PND agrees with the tangent to the congruence. In fact, for each congruence, there is a wide class of NFE solutions specified by the foliation and the value of the field on the field sheet. This is in contrast with null FFE. For null FFE, it has been shown \cite{Menon:2026xnj} that in order for the null geodesic congruence to have a null FFE solution, the foliation needs to be taken so that a component of the shear of the congruence aligned with the field sheet vanishes. For NFE, we do not need any restriction on the shear; thus, NFE may be viewed as a generalization of null FFE in which the restriction on the shear is relaxed. 

Our goal is to construct local NFE solutions from a local null geodesic congruence. Suppose that \(\ell^a\) is tangent to an affinely parametrized null geodesic congruence defined in some local region of spacetime. To construct the field, we need to specify the spacetime foliation. Locally, this can be done as follows. Choose a local spacelike hypersurface $\Sigma$ orthogonal to $\ell$ and pick a spacelike vector field \(\eta^a\) tangent to $\Sigma$. Then, Lie transport $\eta^{a}$ along the congruence. In other words,
\begin{equation}
    \label{eq:eta Lie transport}
    \mathcal{L}_{\ell}\eta=[\ell,\eta]=0.
\end{equation}
Along the congruence, $\eta\cdot\ell=0$ continues to hold since computing the change of $\ell\cdot\eta$ along the geodesic leads to 
\begin{align}
    \ell^{a}\nabla_{a}\left(\ell_{b}\eta^{b}\right)
    &=\ell_{b}\ell_{a}\nabla_{a}\eta^{b}\notag
    \\
    &=\ell_{b}\eta^{a}\nabla_{a}\ell_{b}\notag
    \\
    &=\frac{1}{2}\eta^{a}\nabla_{a} (\ell^{b}\ell_{b})\notag
    \\
    &=0,
\end{align}
where we have used the fact that $\ell^{a}$ is affinely parametrized and tangent to a null geodesic, together with Eq.~\eqref{eq:eta Lie transport} with $[\ell,\eta]^{a}=\ell^{b}\nabla_{b}\eta^{a}-\eta^{b}\nabla_{b}\ell^{a}$.
Equation~\eqref{eq:eta Lie transport} means that the vector space spanned by \(\ell^a\) and \(\eta^a\) is
involutive, so the Frobenius theorem establishes the existence of submanifolds which foliate spacetime locally. Furthermore, $\ell$ is orthogonal to the submanifold since $\ell\cdot\ell=0=\eta\,\cdot\,\ell$, which implies that the submanifolds are null two-surfaces \cite{DuggalBejancu1996}.  
Now, we choose local coordinates
\((\lambda,r,y^1,y^2)\) adapted to the chosen foliation
\begin{equation}
    \ell=\partial_\lambda,\qquad
    \eta=\partial_r .
\end{equation}
Here \(\lambda\) is the affine parameter along the null geodesics, \(r\) is the
second coordinate on each field sheet along $\eta^{a}$, and \(y^1,y^2\) label neighboring field
sheets.

In these coordinates, we may write a closed two-form
\begin{align}
    \label{eq: NFE sol from null congruence}
    F=A(y^{1},y^{2})\,dy^{1}\wedge dy^{2},
\end{align}
where \(A(y^1,y^2)\) is an arbitrary function on the space of field sheets.
This field is indeed closed, \(dF=0\), and degenerate,
\begin{align}
    F\wedge F=0\quad
    \Longrightarrow\quad
    F_{ab}\tilde F^{ab}=0.
\end{align}
Moreover, by construction,
\begin{equation}
    \label{eq: F ell=0=F eta}
    F_{ab}\ell^{b}=0=F_{ab}\eta^{b},
\end{equation}
so the field sheets of the field \eqref{eq: NFE sol from null congruence} are indeed the chosen foliation. 
To show that the constructed $F$ is null, introduce a tetrad adapted to the field sheet.  Let \(s^a=\eta^a/\sqrt{\eta^2}\) be the unit spacelike direction tangent
to the sheet, and complete it to a null tetrad \((\ell,n,\alpha,s)\) satisfying
\begin{align}
    \label{eq: tetrad properties}
    \ell^2=n^2&=0,
    &
    \ell\cdot n&=-1,
    &
    \alpha^2=s^2&=1,
\end{align}
with all other inner products vanishing.  Here \(\alpha^a\) is the spacelike direction orthogonal to the field sheet and $n^{a}$ is the other null direction.  Since \(F\) annihilates
\(\ell\) and \(s\), a direct tetrad calculation shows that it can only contain
the \(\ell_{[a}\alpha_{b]}\) component.\footnote{In the tetrad
\((\ell,n,\alpha,s)\), write the most general
antisymmetric tensor as
\[
\begin{aligned}
F_{ab}
&=
c_1\,\ell_{[a}n_{b]}
  +c_2\,\ell_{[a}\alpha_{b]}
  +c_3\,\ell_{[a}s_{b]}
  +c_4\,n_{[a}\alpha_{b]}  \\
&\quad
+c_5\,n_{[a}s_{b]}
+c_6\,\alpha_{[a}s_{b]} .
\end{aligned}
\]
Imposing \(F_{ab}\ell^b=0\) and \(F_{ab}s^b=0\) gives
\(c_1=c_3=c_4=c_5=c_6=0\), leaving only the
\(c_2\,\ell_{[a}\alpha_{b]}\) term.}  Thus
\(F\) can be written locally as
\begin{align}
    \label{eq:null-field-tetrad-form}
    F_{ab}
    =\Psi(\ell_a\alpha_b-\alpha_a\ell_b)
    =2\Psi\,\ell_{[a}\alpha_{b]},
\end{align}
for some scalar \(\Psi\).  From this expression \(F^{ab}F_{ab}=0\) follows
immediately.  Thus the field is null, and \(\ell^a\) is its PND.
By construction, $\ell$ is tangent to a null geodesic; thus the null-force condition \eqref{eq: null force density} follows immediately from Eq.~\eqref{eq:geodesic to null force}, establishing that a generic null geodesic congruence gives rise to NFE fields. In fact, this construction shows that for a given null geodesic congruence, we can find a large class of NFE solutions specified by the field sheets (i.e., choosing $\eta^{a}$) and the function $A(y^{1},y^{2})$.

Now, let us consider how the Lorentz force density may be written in terms of the shear of the congruence. 
To do so, note that in the tetrad  \((\ell,n,\alpha,s)\),  the metric decomposes as
\begin{align}
    \label{eq: metric null tetrad}
    g_{ab}=-\ell_{a}n_{b}-n_{a}\ell_{b}+Q_{ab},
\end{align}
where
\begin{align}
    \label{eq: define Q}
    Q_{ab}=\alpha_{a}\alpha_{b}+s_{a}s_{b}.
\end{align}
Then, the expansion of the congruence is
(e.g. \cite{Carroll2019})
\begin{align}
    \label{eq: define theta expansion}
    \theta
    =Q^{ab}\nabla_a\ell_b
    =\nabla_a\ell^a,
\end{align}
and its shear is
\begin{align}
    \sigma_{ab}
    &=
    B_{(ab)}-\frac12\theta Q_{ab},
\end{align}
where
\begin{align}
    B_{ab}=Q_a{}^{c}Q_b{}^{d}\nabla_c\ell_d .
\end{align}
The component of the shear along the $s^{a}$ direction will be relevant for the result, so we define
\begin{align}
     \sigma_{ss}&:=s^{a}s^{b}\sigma_{ab}\notag
     \\
     &=s^a s^b\nabla_a\ell_b-\frac12\theta\notag
     \\
     &=\frac{1}{2}\left(s^a s^b- \alpha^a\alpha^b\right)\nabla_a\ell_b.
      \label{eq:sigma-s-s}
\end{align}
Geometrically, this shear component quantifies the extent of compression or stretching of the congruence in the direction tangent to the field sheet. Thus, we refer to $ \sigma_{ss}$ as the shear component aligned with the field sheets.

Now let us compute the Lorentz force density in terms of the shear.  Since \(\ell\cdot j=0\) from
\eqref{eq: j.ell=0}, Eq.~\eqref{eq:null-field-tetrad-form} shows that only the term with 
\(\alpha\cdot j\) survives when $j^{a}$ is dotted into $F_{ab}$.  Recalling \(j^a=\nabla_bF^{ab}\), we have
\begin{align}
    \label{eq:alpha.j calculation}
    \alpha\cdot j
    &=
    \alpha_a\nabla_bF^{ab} \notag\\
    &=
    -\ell^a\nabla_a\Psi
    +\Psi\left(
    \alpha^a\alpha^b\nabla_a\ell_b
    -\nabla_a\ell^a
    \right).
\end{align}
Now consider contracting the homogeneous Maxwell equation \eqref{eq: dF=0} with \(\ell^a n^b\alpha^c\). First note that
\begin{align}
    3\nabla_{[a}F_{bc]}\ell^a n^b\alpha^c
    &=
    \ell^a n^b\alpha^c
    \left(\nabla_aF_{bc}+\nabla_bF_{ca}+\nabla_cF_{ab}\right).
\end{align}
Using \eqref{eq:null-field-tetrad-form}, the three terms on the right-hand
side become
\begin{align}
    \ell^a n^b\alpha^c\nabla_aF_{bc}
    &=
    -\ell^a\nabla_a\Psi,
    \\
    \ell^a n^b\alpha^c\nabla_bF_{ca}
    &=
    0,
    \\
    \ell^a n^b\alpha^c\nabla_cF_{ab}
    &=
    -\Psi\,\alpha^a\alpha^b\nabla_a\ell_b .
\end{align}
Here we have used the tetrad relations \eqref{eq: tetrad properties} and the
affine geodesic condition \eqref{eq: affine null geodesics}.  Therefore, Eq.~\eqref{eq: dF=0} gives
\begin{align}
    \label{eq: dF=0 with congruence}
    0
    &=
    \nabla_{[a}F_{bc]}\ell^a n^b\alpha^c \notag\\
    &=
    -\frac13
    \left(
    \ell^a\nabla_a\Psi
    +\Psi\,\alpha^a\alpha^b\nabla_a\ell_b
    \right),
\end{align}
or equivalently
\begin{align}
    \ell^a\nabla_a\Psi
    =
    -\Psi\,\alpha^a\alpha^b\nabla_a\ell_b .        \label{eq:Psi-transport-shear}
\end{align}
Using \eqref{eq:Psi-transport-shear} in Eq.~\eqref{eq:alpha.j calculation}, and recalling
\(\theta=\nabla_a\ell^a\), gives
\begin{align}
    j_a\alpha^a
    &=
    2\Psi
    \left(
    \alpha^a\alpha^b\nabla_a\ell_b-\frac12\theta
    \right)\notag
    \\
    &= \Psi
    \left(
    \alpha^a\alpha^b\nabla_a\ell_b-s^a s^b\nabla_a\ell_b
    \right)\notag
    \\
    &=
    -2\Psi\sigma_{ss}.                     \label{eq:j-alpha-shear}
\end{align}
Thus, computing $F_{ab}j^b$ using Eq.~\eqref{eq:null-field-tetrad-form} and using Eq.~\eqref{eq:j-alpha-shear}, the Lorentz force density is
\begin{align}
    F_{ab}j^b
    =\Psi (\alpha\cdot j)\ell_{a}
    =
    -2\Psi^2\sigma_{ss}\ell_a .       \label{eq:null-force-shear}
\end{align}
This expression shows the role of the shear in the Lorentz force density, and gives another perspective on the difference between NFE and null FFE. 
Comparing \eqref{eq:null-force-shear} with \eqref{eq:geodesic to null force}, we have 
\begin{align}
    \label{eq: nabla Tem = -shear}
    \nabla_b T^{ab}_{\rm EM}
    =-F\indices{^a_b}j^{b}
    =+2\Psi^2\sigma_{ss}\ell^{a},
\end{align}
Since the total stress tensor $T^{ab}_{\rm NFE}=T_{\rm EM}^{ab}+T_{\rm m}^{ab}$ is conserved \eqref{eq: NFE energy conservation}, we also have
\begin{align}
    \label{eq: nabla Tm = +shear}
    \nabla_b T^{ab}_{\rm m}
    &=-2\Psi^2\sigma_{ss}\ell^{a}.
\end{align}
The field-aligned shear \(\sigma_{ss}\) therefore controls the exchange
of stress-energy between the electromagnetic field and matter.  When
\(\sigma_{ss}=0\), the electromagnetic stress tensor is conserved by
itself and the null FFE is recovered.  When
\(\sigma_{ss}\neq0\), the electromagnetic field exchanges
energy-momentum with the null dust, and the system is genuinely NFE.

\subsection{Plasma energy in NFE region}
A special feature of NFE is that the matter stress-energy tensor can be determined from the electromagnetic field once the total NFE field is known, for example, using Eq.~\eqref{eq: null force matter} to find $\phim$ from the field. 
Here, we show a more explicit method to find $\phim$, and subsequently $T_{\rm m}^{ab}$ (Eq.~\eqref{eq: T_matter null dust} connects $\phim$ and $T_{\rm m}^{ab}$),  using the properties of null congruence.
Since \(\ell^a\) is affinely parametrized, Eq.~\eqref{eq: affine null geodesics} removes the first term in Eq.~\eqref{eq: NFE energy conservation}. Since $\ell^{b}$ is nonzero, from Eq.~\eqref{eq: NFE energy conservation}, we have $\nabla_{a}(\varphi_{\rm NFE}\ell^{a})=0$. 
Let \(\lambda\) be an affine parameter along a geodesic.  Then using
\(\ell^a\nabla_a=\dd/\dd\lambda\), we find
\begin{align}
    0=\nabla_a(\varphi_{\rm NFE}\ell^a)
    &=
    \frac{\dd\varphi_{\rm NFE}}{\dd\lambda}
    +\varphi_{\rm NFE}\,\nabla_a\ell^a . 
\end{align}
Equivalently,
\begin{align}
    \frac{\dd}{\dd\lambda}\ln\varphi_{\rm NFE}
    &=-\theta.
\end{align}
where $\theta$ is the expansion of the congruence \eqref{eq: define theta expansion}. 
Therefore, if \(\lambda=0\) is the point where the NFE region emerges,
\begin{align}
    \varphi_{\rm NFE}(\lambda)
    &=
    \varphi_{\rm NFE}(0)
    \exp\left[
    -\int_0^\lambda \theta(\bar\lambda)\,\dd\bar\lambda
    \right].                                      \label{eq:phi-general-integral}
\end{align}
The matter energy density is then obtained by subtracting the electromagnetic contribution:
\begin{align}
    \label{eq: NFE matter energy}
    \varphi_{\rm m}(\lambda)
    &=\varphi_{\rm NFE}(\lambda)-\varphi_{\rm EM}(\lambda). 
\end{align}
This procedure is local along each geodesic: compute the expansion $\theta$ of the null congruence, integrate $\theta$ along it and subtract the electromagnetic energy density determined from Eq.~\eqref{eq: T_EM}. 
The matter stress tensor then is obtained from Eq.~\eqref{eq: T_matter null dust}.

\subsection{Initial value formulation}\label{sec: initial value formulation}
Now that we have established the connection between NFE and null geodesic congruences, let us turn to the initial value formulation of NFE and the local construction of solutions. 
Let \(\Sigma\) be a spacelike hypersurface embedded in a spacetime $M$,
\begin{align}
    \iota:\Sigma\rightarrow M,
\end{align}
and choose local coordinates \(z^i\) on \(\Sigma\). The initial data are a nonzero null spacetime two-form \(F_0\) restricted to \(\Sigma\), together with its PND \(\ell_0^a\), satisfying
\begin{align}
    \label{eq: PND at Sigma}
    (F_0)_{ab}\ell_0^b=0 .
\end{align}
Let $f$ be the intrinsic two-form defined by the pullback of $F_{0}$ to $\Sigma$:
\begin{equation}
    \label{eq: define f}
    f:=\iota^*F_0
      =\frac12 f_{ij}(z)\,dz^i\wedge dz^j,
\end{equation}
where $f_{ij}(z)$ are the components of $f$ in the $z^{i}$ coordinates on $\Sigma$. 

The procedure for constructing the field away from $\Sigma$ starts by launching the affinely parametrized null geodesic from $z\in \Sigma$ with initial tangent \(\ell_0^a(z)\). This defines the ray map
\begin{align}
    \label{eq: general ray map}
    x^\mu=\mathcal{X}^{\mu}(\lambda,z),
\end{align}
with initial conditions
\begin{align}
    \mathcal{X}^{\mu}(0,z)&=\iota^{\mu}(z),
    \\
    \frac{\partial\mathcal{X}^{\mu}}{\partial\lambda}(0,z)
    &=\ell_0^{\mu}(z),
\end{align}
where $x^{\mu}$ are the spacetime coordinates. 
Before conjugate points form, $(\lambda,z^i)$ are good
coordinates on the NFE region swept out by the congruence.
In these ray-adapted coordinates,
\begin{equation}
    \label{eq: ell =del lambda}
    \ell=\partial_\lambda.
\end{equation}
Equation~\eqref{eq: PND at Sigma} together with Eq.~\eqref{eq: ell =del lambda} forces the coefficients of $d\lambda\,\wedge\, dz^i$ in $F_{0}$ to vanish. This means that only the $dz^i\wedge dz^j$ components of $F_{0}$ are nonzero, but these components are equal to the components of $f$ since the pullback gives $\iota^*d\lambda=0$ and $\iota^*dz^{i}=dz^{i}$.  Therefore, $F_{0}$ in the ray-adapted coordinates is written as
\begin{equation}
    \label{eq: F0 on Sigma}
    F_0=\frac12 f_{ij}(z)\,dz^i\wedge dz^j
    \qquad \text{on } \Sigma .
\end{equation}
Now, consider finding an NFE solution away from $\Sigma$ with the data given by Eq.~\eqref{eq: F0 on Sigma}. To find the solution, we use the fact that $F_{ab}\ell^{b}=0$ holds away from $\Sigma$. Using the same argument as in the derivation of Eq.~\eqref{eq: F0 on Sigma}, one can show that the $d\lambda\wedge dz^i$ components vanish in these coordinates. Hence the field away from $\Sigma$ has the general form
\begin{align}
    \label{eq: general F in geodesic coord}
    F
    =
    \frac12 F_{ij}(\lambda,z)\,\dd z^i\wedge\dd z^j.
\end{align}
The homogeneous Maxwell equation \eqref{eq: dF=0} then gives
\begin{align}
    0=(\dd F)_{\lambda ij}=\partial_\lambda F_{ij},
\end{align}
so the components \(F_{ij}\) are constant along each ray. At \(\lambda=0\), $F_{ij}$ needs to agree with $(F_{0})_{ij}$, which in turn implies that it agrees with the pullback data:
\begin{align}
    F_{ij}(0,z)=(F_{0})_{ij}(z)=f_{ij}(z).
\end{align}
Let \(z^i=Z^i(x)\) denote the ray label obtained by inverting Eq.~\eqref{eq: general ray map}. Then the NFE field at a point $x$ is constructed as 
\begin{align}
    F
    &=\frac{1}{2}f_{ij}(Z(x))dZ^{i}\wedge dZ^{j}\notag
    \\
    &=
    \frac12 f_{ij}(Z(x))\,\frac{\partial Z^{i}(x)}{\partial x^{\mu}}\frac{\partial Z^{j}(x)}{\partial x^{\nu}}d x^{\mu}\wedge dx^{\nu}.       \label{eq:nfe-label-field}
\end{align}
This expression agrees with $F_{0}$ given by Eq.~\eqref{eq: F0 on Sigma} at $\lambda=0$. 
Also, it manifestly satisfies \eqref{eq: dF=0} and \eqref{eq: Fell=0}. This gives a practical construction procedure: first, compute the pullback \(f=\iota^*F_0\) of the initial data $F_{0}$ on $\Sigma$, launch the null geodesic congruence from \(\Sigma\) in the direction of $\ell_{0}$, and evaluate Eq.~\eqref{eq:nfe-label-field} using the ray map \(Z^i(x)\). 
This is the sense in which the solution is obtained by the method of characteristics: our field is found by first solving the null geodesic equation (finding characteristics), and then the NFE field is obtained by transporting the initial data along the characteristics. 

An equivalent but more geometrical perspective on this construction is also possible. First, define a map \(\mathscr{P}:U\to\Sigma\) that sends each spacetime point to a point $z$ on $\Sigma$ along the null geodesic whose tangent agrees with $\ell_{0}$ on $\Sigma$.  Then, Eq.~\eqref{eq:nfe-label-field} simply states that $F$ is the pullback of $f$ by the map $\mathscr{P}$. Using Eq.~\eqref{eq: define f}, the constructed field is in the compact form:
\begin{align}
    F=\mathscr{P}^*(\iota^{*}F_{0}).
\end{align}
The transport property of NFE is seen in Eq.~\eqref{eq: Lie F =0}. The construction we have shown here is a reflection of this property.

\subsection{\texorpdfstring{$3+1$}{3+1} form of NFE and Ohm's law}
Just as in the case of FFE, NFE can be written in 3+1 form. To see this, let us look at FFE in 3+1 form first. 
Using the electric field $\vec{E}$ and magnetic field $\vec{B}$, Maxwell's equations are
\begin{align}
    \label{eq: 3+1D maxwell1}
    \partial_t \vec B+\nabla\times \vec E&=0,
    \\
    \label{eq: 3+1D maxwell2}
    \partial_t \vec E-\nabla\times \vec B&=-\vec j,
\end{align}
with the constraints $\nabla\cdot \vec B=0$ and $\nabla\cdot \vec E=\rho$, and 
\(\rho\) and \(\vec j\) are the charge and current densities.
In this case, the force-free condition \eqref{eq: force free condition} is given by
\begin{align}
    \vec{j}\cdot \vec{E}=0,\qquad \rho \vec{E}+\vec{j}\times \vec{B}=0.
    \label{eq: 3+1 force free condition}
\end{align}
The first condition is that $\vec{j}$ has no $\vec{E}$ component. From the second one, the $\vec{E}\times\vec{B}$ component of $\vec{j}$ is found to be $\vec{j}_{\perp}=|\vec{B}|^{-2} \rho\,\vec{E}\times\vec{B}$ by taking $\times \vec{B}$ of Eq.~\eqref{eq: 3+1 force free condition}. 
The component of $\vec{j}$ that is parallel to $\vec{B}$ can also be found \cite{gruzinov1999stabilityforcefreeelectrodynamics} by taking the time derivative of $\vec{E}\cdot \vec{B}=0$, and using Eqs.~\eqref{eq: 3+1D maxwell1} and \eqref{eq: 3+1D maxwell2}. 
This results in the expression of $\vec{j}$ solely in terms of the field and its spatial derivative:
\begin{align}
    \label{eq: FFE Ohm's law}
    &\vec{j}_{\rm FFE}\notag
    \\&=
    \frac{(\nabla\cdot \vec E)(\vec E\times \vec B)+\vec B\left[
    \vec B\cdot (\nabla\times \vec B)
    -\vec E\cdot(\nabla\times \vec E)
    \right]}{\vec B^{\,2}} .
\end{align}
This ``Ohm's law'' for FFE can then be plugged back into Eq.~\eqref{eq: 3+1D maxwell2} to solve for the time evolution of $\vec{E}$ and $\vec{B}$. 
This formulation is hyperbolic \cite{Komissarov2002nov,Pfeiffer:2013wza,Palenzuela:2011es,Carrasco_2016} as long as $\vec{B}^{\,2}-\vec{E}^{\,2}$ is positive. 

The same process can be used to derive a similar Ohm's law for NFE. 
NFE requires the field to be null \eqref{eq: null electromagnetic field}, which in \(3+1\) form is given by
\begin{align}
    \label{eq: 3+1 null field condition}
    \vec E\cdot \vec B=0,
    \qquad
    \vec B^{\,2}-\vec E^{\,2}=0.
\end{align}
In flat spacetime, the affinely parametrized PND may be written as
\begin{align}
    \label{eq: PND flat spacetime}
    \ell^{\mu}=(1,\vec{\ell}),
\end{align}
where the spatial component \(\vec{\ell}\) is given by
\begin{align}
    \label{eq: PND flat spacetime spatial}
    \vec{\ell}:=\frac{\vec E\times \vec B}{\vec B^{\,2}}.
\end{align}
The spatial part of the null-force condition \eqref{eq: null force density} in 3+1 form is given by
\begin{align}
    \rho \vec{E}+\vec{j}\times\vec{B}\propto\vec{\ell}.
\end{align}
By explicitly writing the proportionality constant $C$ and using the time component of Eq.~\eqref{eq: null force density}, we have
\begin{align}
    \label{eq: 3+1 null force time}
    \vec{j}\cdot\vec{E}&=C,
    \\
    \label{eq: 3+1 null force spatial}
    \rho \vec{E}+\vec{j}\times\vec{B}&=C\vec{\ell}.
\end{align}
By eliminating $C$ from Eq.~\eqref{eq: 3+1 null force spatial} using Eq.~\eqref{eq: 3+1 null force time} and using Eq.~\eqref{eq: PND flat spacetime spatial} to explicitly write $\vec{\ell}$, we obtain the field-only form of the null-force condition
\begin{align}
    \label{eq: 3+1 null force field only}
    \rho \vec{E}+\vec{j}\times\vec{B}&=(\vec{j}\cdot\vec{E})\frac{\vec E\times \vec B}{\vec B^{\,2}}.
\end{align}
Furthermore, dotting $\vec{j}$ into Eq.~\eqref{eq: 3+1 null force field only}, we obtain
\begin{align}
    \label{eq: 3+1 NFE condition}
    \vec{\ell}\cdot\vec j=\rho.
\end{align}
This is a 3+1 form of $\ell\cdot j=0$ derived in Eq.~\eqref{eq: j.ell=0}. 
Equation~\eqref{eq: 3+1 NFE condition} fixes the component of the current along the null direction.
To obtain a closed \(3+1\) expression for \(\vec j\), we also require preservation of the null constraints in time. Differentiating \(\vec E\cdot \vec B=0\) and \(\vec B^{\,2}-\vec E^{\,2}=0\) in time and using Maxwell's equations yields
\begin{align}
    \label{eq: 3+1 J B component}
    \vec B\cdot \vec j
    &=
    \vec B\cdot (\nabla\times \vec B)
    -\vec E\cdot(\nabla\times \vec E),
    \\
    \label{eq: 3+1 J E component}
    \vec E\cdot \vec j
    &=
    \vec B\cdot(\nabla\times \vec E)
    +\vec E\cdot(\nabla\times \vec B).
\end{align}
Since \(\{\vec E,\vec B,\vec{\ell}\}\) form an orthogonal basis when
\(\vec B^{\,2}=\vec E^{\,2}\neq 0\), the current decomposes as
\begin{align}
    \vec j=\alpha \vec E+\beta \vec B+\rho \vec{\ell},
\end{align}
and all the components are fixed by Eqs.~\eqref{eq: 3+1 NFE condition}--\eqref{eq: 3+1 J E component}.
Therefore NFE Ohm's law may be written as
\begin{align}
    \label{eq: NFE Ohm 3+1}
    \vec j_{\rm NFE}\notag
    &=\frac{\vec B\left[
    \vec B\cdot (\nabla\times \vec B)
    -\vec E\cdot(\nabla\times \vec E)
    \right]}{\vec B^{\,2}}\notag
    \\
    &\hspace{0.5cm} 
    +\frac{\nabla\cdot\vec{E}(\vec{E}\times\vec{B})}{\vec{B}^{2}}\notag
    \\
    &\hspace{1cm}+\frac{
     \vec B\cdot(\nabla\times \vec E)
    +\vec E\cdot(\nabla\times \vec B)}
    {\vec B^{\,2}}\vec E\notag
    \\
    &=\vec{j}_{\rm FFE}
    +\frac{ \vec B\cdot(\nabla\times \vec E)
    +\vec E\cdot(\nabla\times \vec B)}
    {\vec B^{\,2}}\vec E,
\end{align}
where \(\vec{j}_{\rm FFE}\) is the current given by FFE Ohm's law \eqref{eq: FFE Ohm's law}.

Equation~(\ref{eq: NFE Ohm 3+1}) is the direct analogue, for NFE, of the familiar \(3+1\) current closure for FFE. The main difference is that for NFE, \(\vec j\cdot\vec E\neq0\), implying that the energy of the EM field alone is not conserved, as we have seen (e.g. Eq.~\eqref{eq:geodesic to null force}). Thus, the 3+1 form of Ohm's law for NFE provides a clear view of the differences and similarities between FFE and NFE. 
Unlike FFE, however, NFE in this 3+1 form is \textit{not} well-posed as an initial value problem by itself. 
The linear stability analysis is given in Appendix~\ref{app:nfe-3plus1-illposedness}, showing explicitly that finer perturbations grow increasingly faster.
Nevertheless, this should not be interpreted as a failure of NFE; rather, it simply means that this particular evolution form is not suitable and the right formulation needs to be based on the transport approach.

\section{Current sheet and FFE/NFE interface}\label{sec:current sheet interface}
In the previous section, we discussed the general properties of NFE. In this section, we focus on the current sheet and the FFE/NFE interface. The description of these objects require additional physical input that goes beyond the smooth NFE description. However, as we will show in Sec.~\ref{sec:applications}, they play an important role in constructing a self-consistent evolution of the field in a physically relevant situations.  Following the spirit of describing plasma physics as much as possible without plasma, we do not attempt to model the details of the current sheet and FFE/NFE interface. Instead, our treatment aims for maximally plasma-free description while still capturing the essential features of these objects.

\subsection{Formation of current sheets}\label{subsec: current sheet}
As shown in Sec.~\ref{sec: initial value formulation}, the NFE field is carried along the null congruence. However, a generic null congruence can focus and form conjugate points in finite time \cite{Wald:1984rg,Hawking:1973uf,Poisson:2009pwt}. At such a point, the smooth evolution derived in Sec.~\ref{sec: initial value formulation} stops since the inversion of the ray map becomes impossible.  The natural continuation is then not a smooth evolution but a discontinuous one, together with an appropriate rule for the field outside the discontinuity. The singularity formation due to conjugate-point formation of the null congruence is in fact analogous to shock formation by ``characteristic crossing'' in nonlinear partial differential equations (PDEs), with the inviscid Burgers equation as the canonical example \cite{Laxbook,evans10}.
In the electromagnetic context, the loss of smooth evolution and subsequent appearance of discontinuity in the field signals the formation of a \textit{current sheet} \cite{Ripperda_2021,N_ttil__2022,Li_Xinyu_Beloborodov_2021}.

A physical current sheet is a thin kinetic layer and the details of microscopic plasma physics are required to model its dynamics \cite{Sironi_2014,Cerutti_2015,Sironi:2025kgn}. 
The basis of our treatment of the current sheet relies on \cite{Li_Xinyu_Beloborodov_2021}, where it is treated as a boundary condition motivated by the plasma screening of the electric field inside the current sheet. We adopt this perspective, and use the covariantly generalized version of their boundary as the model for our current sheet. Our consideration is also limited to a particular class of sheet geometries. For the Alfv\'en-wave collision example we will discuss in Sec.~\ref{subsec: Alfven wave collision}, the sheet appears as a timelike hypersurface. Thus, we focus only on a sheet with this geometry and leave other geometries for future work.

Following \cite{Gralla.Jacobson2014}, let $s$ be a smooth scalar function that vanishes on a timelike hypersurface $\mathcal{C}$ representing the current sheet. We choose $s$ to increase smoothly when moving from one side of the sheet to the other. Across the sheet, the field can be written as
\begin{align}
    \label{eq: F=F+ + F- current sheet}
    F=F^{+}\Theta(s)+F^{-}\Theta(-s),
\end{align}
where $F^{+}$ and $F^{-}$ denote the field on each side. Each side satisfies $dF^{\pm}=0$, and requiring the full distributional field to satisfy $dF=0$ gives
\begin{align}
    0=dF=\delta(s)ds\wedge [F],
\end{align}
where $[F]=F^{+}-F^{-}$ evaluated on $\mathcal{C}$.
Thus, on $\mathcal{C}$,
\begin{align}
    ds\wedge [F]=0.
\end{align}
Equivalently, the pullback $\mathcal{F}_{\mathcal{C}}:=\iota_{\mathcal{C}}^*F$ of the field tensor to $\mathcal{C}$ (where $\iota:\mathcal{C}\to M$ is a map) is continuous:
\begin{align}
    \label{eq: pullback continuity current sheet}
    [\mathcal{F}_{\mathcal{C}}]=0.
\end{align}
Thus the pullback of $F$ to the sheet is continuous.

For a charged particle with velocity $v^{a}$, charge $q$, and mass $m$, the equation of motion is  $mv^{b}\nabla_{b}v^{a}=qF\indices{^a_b}v^{b}$. If the particle is perfectly contained in a sheet (i.e., $v$ is tangent to $\mathcal{C}$), then the pullback of the equation of motion gives $mv^{B}D_{B}v^{A}=q(\mathcal{F}_{\mathcal{C}})\indices{^A_B}v^{B}$, where indices $A,B,\dots$ indicate tensor indices intrinsic to $\mathcal{C}$, and $D_{A}$ is a current-sheet intrinsic covariant derivative.
Thus, the pullback $\mathcal{F}_{\mathcal{C}}$ determines the force experienced by charges constrained to move within the sheet. It is therefore natural to classify it as magnetically dominated, electrically dominated, or null with respect to the induced metric on $\mathcal{C}$. If the intrinsic field were electrically dominated, charged particles in the sheet would be strongly accelerated and would tend to screen that electric field. We therefore impose the analogue of the NFE null condition on the sheet:
\begin{align}
    \label{eq: null CS}
    (\mathcal{F}_{\mathcal{C}})^{AB}(\mathcal{F}_{\mathcal{C}})_{AB}=0.
\end{align}
This is our boundary condition for the sheet. This prescription is intended as an effective continuation after discontinuity formation, and is not meant to capture everything about the current sheet. For instance, our prescription does not say anything about the lifetime of the current sheet. Once the energy supply to the current sheet is stopped, a realistic current sheet dissipates energy through various mechanisms and the sheet eventually disappears. Instead of capturing this effect, the boundary-condition prescription is meant to work within some specified interval after the sheet formation during which the macroscopic description applies. 

Using Maxwell's equation \eqref{eq: nabla F=j} and Eq.~\eqref{eq: F=F+ + F- current sheet}, the current is given by
\begin{align}
    j^{a}
    &=j_{+}^{a}\Theta(s)+j_{-}^{a}\Theta(-s)+\delta(s)(ds)_{b}[F^{ab}],
\end{align}
where $j_{\pm}^{a}=\nabla_{b}F^{ab}_{\pm}$ are bulk currents. Thus the corresponding surface current is
\begin{align}
    \label{eq: current sheet current vector}
     j^a_{\mathcal{C}}=\delta(s)(ds)_b[F^{ab}].
\end{align}
This implies 
\begin{align}
    (ds)_{a}j^{a}_{\mathcal{C}}=0.
\end{align}
Since $ds$ is orthogonal to $\mathcal{C}$, this equation means that $j_{\mathcal{C}}^{a}$ is tangent to $\mathcal{C}$.

\subsection{FFE/NFE interface}
\label{subsec:junction condition}
For astrophysically important situations involving the breakdown of FFE (e.g. current sheets in pulsars \cite{Spitkovsky_2006,Uzdensky_2013,Cerutti_2015}, collisions of Alfv\'en waves \cite{Li_Xinyu_Beloborodov_2021,N_ttil__2022,Ripperda_2021}), the transition from FFE to NFE happens in some regions of spacetime while other regions remain FFE. 
This means that the FFE and NFE regions must coexist in spacetime, and their boundary is dynamical. 
Our task here is to consider general properties of the FFE/NFE interface.  

Let us define $\Omega_{\rm FFE}$ and $\Omega_{\rm NFE}$ as the spacetime regions governed by FFE and NFE, respectively. 
We assume that the spacetime is initially governed entirely by FFE and that $\Omega_{\rm NFE}$ forms only later, when magnetic dominance is lost in FFE. 
This observation gives rise to two notions of FFE/NFE boundaries. The first kind is the \textit{birth front} $\Birth$ of NFE.  This is a region of spacetime where the hyperbolicity of FFE evolution ends, and the FFE data are transferred as the initial data for NFE, which are then Lie transported along null geodesics to find the solution in $\Omega_{\rm NFE}$. Thus, the field is \textit{continuous} across $\Birth$. To ensure that the data on $\Birth$ are fully specified from the FFE region, we assume that $\Birth$ is a spacelike hypersurface \footnote{As far as we know, there is no theorem stating that the null region formed from the FFE evolution must be spacelike. However, for the examples used in this paper, $\Birth$ is indeed spacelike, so our assumption is applicable.}. This way, we may clearly separate the region with and without $\Omega_{\rm NFE}$: in the past of $\Birth$ (i.e. spacetime points reached by timelike curves by following the curve backward in time from $\Birth$), $\Omega_{\rm NFE}$ does not exist yet, and only in the future of $\Birth$, $\Omega_{\rm NFE}$ appears.

\begin{figure}[ht]
\centering
\begin{tikzpicture}[
  interface/.style={very thick, line cap=round, line join=round}
]

\draw[interface]
  (-1.95,3.25) .. controls (-1.95,2.20) and (-2.05,0.80) .. (-2.20,0)
  .. controls (-0.90,-0.42) and (0.90,-0.42) .. (2.20,0)
  .. controls (2.05,0.80) and (1.85,2.20) .. (1.75,3.25);

\node[font=\large] at (0,1.3) {$\Omega_{\mathrm{NFE}}$};
\node[font=\large] at (0,-1.5) {$\Omega_{\mathrm{FFE}}$};

\node[font=\fontsize{14}{16}\selectfont] at (-2.5,1.55) {$I$};
\node[font=\fontsize{14}{16}\selectfont]at (2.5,1.55) {$I$};
\node[font=\fontsize{14}{16}\selectfont,below right] at (0,-0.5) {$\Birth$};

\end{tikzpicture}
\caption{Schematic geometry of the spacelike birth front $\Birth$ and timelike FFE/NFE interface $I$. The data for the NFE region are prescribed on $\Birth$ and are carried along the null geodesics launched from $\Birth$ in the direction of $\ell^{a}$.}
\label{fig:ffe-nfe-interface}
\end{figure}
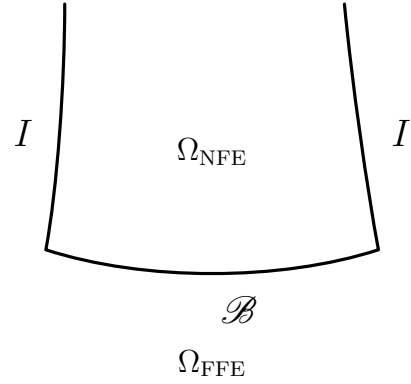

The second type of boundary appears in the future of $\Birth$, where $\Omega_{\rm FFE}$ and $\Omega_{\rm NFE}$ meet. We call this boundary the \textit{FFE/NFE interface} $I$ and assume that it is a timelike hypersurface. Thus, a generic timelike observer sees the interface as a moving layer separating two regions (see Fig.~\ref{fig:ffe-nfe-interface} for the geometry of the birth front $\Birth$ and the FFE/NFE interface $I$).  

We show here that the conditions for the interface $I$ share some properties with the timelike current sheet $\mathcal{C}$ discussed in Sec.~\ref{subsec: current sheet}. Let $\Phi$ be a smooth scalar function that vanishes on the interface and increases when moving from the NFE side to the FFE side. We scale $\Phi$ so that the surface normal $m^{a}=\nabla^{a}\Phi$ has unit length ($m^{a}m_{a}=1$). 
Across the interface, $F$ can be written as
\begin{align}
    F=F_{\rm FFE}\Theta(\Phi)+F_{\rm NFE}\Theta(-\Phi),
\end{align}
Following the same argument used to derive Eq.~\eqref{eq: pullback continuity current sheet}, we find
\begin{align}
    \label{eq: interface pullback wedge}
    d\Phi\wedge[F]=0,
\end{align}
or equivalently
\begin{align}
    \label{eq: interface pullback continuity general}
    [\mathcal{F}_{I}]=0,
\end{align}
where $\mathcal{F}_{I}$ is a pullback of $F$ to the interface $I$, and $[F]=F_{\rm FFE}-F_{\rm NFE}$ is evaluated on $I$.

To fully determine the dynamics of the interface, an additional condition is needed. 
Here we show a no-go theorem for a stress-free interface: in general, the interface must carry its own surface stress. 
To show this, assume that the unresolved interface carries no surface stress-energy. In this case, the Rankine--Hugoniot condition
\cite{SmollerTemple1993sep,Israel1966jul} holds across $I$,
\begin{equation}
    m_a[T^{ab}]=0,
    \label{eq:full-stress-matching-interface}
\end{equation}
We also assume $m\cdot \ell\neq0$, which implies that the interface and the null geodesic congruence intersect. 
Then Eq.~\eqref{eq:full-stress-matching-interface} gives
\begin{equation}
    m_aT_{\rm FFE}^{ab}
    =m_{a}T_{\rm NFE}^{ab}=
    \varphi_{\rm NFE}(m\cdot \ell)\ell^b .
\end{equation}
Here, for the right-hand side, we have used Eq.~\eqref{eq: NFE stress energy tensor}. 
Thus the vector \(t^b:=m_aT_{\rm FFE}^{ab}\) is null.  For the electromagnetic stress tensor \(T_{\rm EM}\) given by Eq.~\eqref{eq: T_EM}, the following algebraic identity holds \cite{Martin-Moruno:2017iqw}:
\begin{align}
    T_{\rm EM}^{a}{}_{c}T_{\rm EM}^{c}{}_{b}
    &=
    \frac{1}{4}\left(T_{\rm EM}^{d}{}_{c}T_{\rm EM}^{c}{}_{d}\right)\delta\indices{^{a}_{b}}\notag
    \\
    &=
    \frac{1}{16}\left((F_{ab}F^{ab})^{2}+(F_{ab}\tilde F^{ab})^{2}\right)\delta^a{}_b.
\end{align}
This, together with $t^{a}$ being null, leads to
\begin{align}
    0=t_b t^b
    &=
    m_aT_{\rm FFE}^{a}{}_{b}
    T_{\rm FFE}^{b}{}_{c}m^c \notag
    \\
    &=
    \frac{1}{16}\left((F_{ab}^{\rm FFE}F^{ab}_{\rm FFE})^{2}
    +(F_{ab}^{\rm FFE}\tilde F^{ab}_{\rm FFE})^{2}\right)m^{a}m_{a},
\end{align}
Since \(m^a m_a=1\), this equation holds only when $F_{ab}^{\rm FFE}F^{ab}_{\rm FFE}=0=F_{ab}^{\rm FFE}\tilde F^{ab}_{\rm FFE}$. Therefore the FFE-side field
must also be null.  This implies that, on the interface, \(T_{\rm FFE}^{ab}\)
can be written as \(T_{\rm FFE}^{ab}=h k^{a}k^{b}\), where \(k^a\) is its own
PND.  Equation~\eqref{eq:full-stress-matching-interface}
then implies
\begin{align}
      h(m\cdot k)k^b=\varphi_{\rm NFE}(m\cdot \ell)\ell^b .
\end{align} 
Assuming that \(h\), \(\varphi_{\rm NFE}\), \(m\cdot k\), and
\(m\cdot \ell\) are nonzero, this forces \(k^a\) to be parallel to
\(\ell^a\).  We use the scaling freedom to set \(k^a=\ell^a\).

Now use the pullback condition \eqref{eq: interface pullback continuity general}. Its equivalent form \eqref{eq: interface pullback wedge} implies
\begin{equation}
    [F]=m\wedge p
\end{equation}
for some one-form \(p\) that is tangent to the interface.  Since the two fields are
null with the same PND,
\begin{equation}
    0=\ell_{a} [F^{ab}]
    =(m\cdot\ell)p^{b}-(p\cdot \ell)m^{b} .
\end{equation}
Provided that \(m\cdot\ell\neq0\), this equation implies \(p=0\), and hence
\begin{equation}
    \label{eq: F continuous sheet}
    [F]=0 .
\end{equation}
Since the EM stress tensor is given by Eq.~\eqref{eq: T_EM} and depends only on $F$, Eq.~\eqref{eq: F continuous sheet} implies that the EM part of the stress-energy tensor is continuous across the interface. 
Then, using the condition~\eqref{eq:full-stress-matching-interface} together with Eq.~\eqref{eq: T_matter null dust}, we have
\begin{align}
    \label{eq: phim=0 RH condition}
    0&=m_{a} [T^{ab}]\notag
    \\
    &=m_{a}\left(T_{\rm FFE}^{ab}-T_{\rm NFE}^{ab}\right)\notag
    \\
    &=-m_{a}T^{ab}_{\rm m}\notag
    \\
    &=-\phim (m \cdot \ell)\ell^{b}.
\end{align}
Since we assumed nonzero $\ell^{a}$ and $m \cdot \ell$, Eq.~\eqref{eq: phim=0 RH condition} implies that the matter contribution \(\varphi_{\rm m}\) on the NFE side must vanish. 
However, this is not compatible with a generic NFE solution. 
The NFE field is determined by the data on $\Birth$ and transported along the null congruence. In other words, the NFE field is independent of the location of the interface, so $\phim$ has no reason to vanish on $I$. Therefore, the condition \eqref{eq:full-stress-matching-interface} does not serve as a generic junction condition for the FFE/NFE interface. 

The implication is that the FFE/NFE interface is generally a surface-energy-carrying layer, i.e.,
\(m_a[T^{ab}]\neq0\).  
To conclude this section, we give a brief overview of how the surface stress-energy enters the theory. 
Let \(S^{AB}\) be an intrinsic surface stress tensor, where $A,B$ are tensor indices intrinsic to the interface.
Let \(y^A\) be
coordinates on the interface, \(e_A{}^a=\partial x^a/\partial y^A\) the basis, \(D_A\) the intrinsic covariant derivative on $I$, and
\(K_{AB}=e_A{}^a e_B{}^b\nabla_a m_b\) the extrinsic curvature.  We write the total stress tensor including the surface stress-energy as \cite{Poisson:2009pwt}
\begin{align}
    T^{ab}
    =
    T_{\rm FFE}^{ab}\Theta(\Phi)
    +T_{\rm NFE}^{ab}\Theta(-\Phi)
    +S^{AB}e_A{}^a e_B{}^b\delta(\Phi).
\end{align}
Conservation, \(\nabla_a T^{ab}=0\), then gives the
surface balance law \cite{BarrabesIsrael1991feb} (see Appendix~\ref{app:surface-balance-law} for derivation):
\begin{align}
    \label{eq: surface stress S eq1}
    D_{A}S^{AB}&=-e\indices{^B_b} m_{a}[T^{ab}],
    \\
    \label{eq: surface stress S eq2}
    K_{AB}S^{AB}&=m_{a}m_{b}[T^{ab}].
\end{align}
Thus the detailed interface dynamics is encoded in \(S^{AB}\).  
We do not attempt to model $S^{AB}$ since it would generally require introducing additional parameters to the theory. For the FFE+NFE application examples used in this paper, we consider an ad hoc model
to close the system.

For a generic $S^{AB}$, there is no reason for Eq.~\eqref{eq: F continuous sheet} to hold; thus, $F$ is generally \textit{discontinuous} across the interface. When $F$ is discontinuous across the interface, the same argument used for Eq.~\eqref{eq: current sheet current vector} leads to the surface current
\begin{align}
    \label{eq: interface current vector}
     j^a_{I}=\delta(\Phi)(d\Phi)_b[F^{ab}].
\end{align}

\section{Application of FFE+NFE theory}\label{sec:applications}
The previous sections explored the general properties of NFE, and we derived various results.  This section demonstrates the application of the NFE continuation from FFE. In doing so, we use two analytic examples of FFE solutions that exhibit the transition from magnetically dominated to electrically dominated in finite time.
The first is the collision of planar, counter-polarized Alfv\'en waves. 
This example provides a full application of the techniques developed above: the birth of an NFE region in an otherwise FFE spacetime, transport along the null congruence, the formation of a current sheet, and the evolution of FFE/NFE interfaces.
The second is Adhikari's temporally type-changing force-free solution \cite{Adhikari:2025rya}. This example is mathematically cleaner: no FFE/NFE interface is formed since the loss of magnetic dominance occurs on an entire Cauchy surface, and the NFE continuation never develops a current sheet; thus the full continuation can be obtained without introducing a current-sheet or interface prescription.

\subsection{Collision of Alfv\'en waves}\label{subsec: Alfven wave collision}
Here we study a collision of two counter-propagating Alfv\'en waves \cite{Li_Xinyu_Beloborodov_2021,Ripperda_2021,N_ttil__2022,Li:2018kco}. 
We follow the setting used in \cite{Li_Xinyu_Beloborodov_2021}, where the Alfv\'en waves are planar-symmetric, counter-polarized, and propagate only in one spatial direction with a background guide magnetic field. 
By counter-polarized, we mean that the two incoming waves have opposite transverse magnetic fields while the electric fields point in the same direction. 
Their collision weakens the transverse magnetic field while strengthening the electric field, decreasing $P=\vec{B}^{2}-\vec{E}^{2}$. We also consider a triangular wave as the initial profile instead of the square-shaped incoming waves considered in \cite{Li_Xinyu_Beloborodov_2021}. 
Under this setup, $P$ gradually reaches zero for sufficiently large amplitude, and an NFE region eventually appears, enabling us to make use of the techniques developed in the previous sections.

We work in flat spacetime and use the usual coordinates $(t,x,y,z)$.  
We assume that the waves propagate in the $z$ direction.  
In the planar-symmetric system, Maxwell's equations are
\begin{align}
    \partial_t B_y&=-\partial_z E_x,
    &
    \partial_t E_x&=-\partial_z B_y-j_x,\label{eq: relevant 1+1D equation}
    \\
    \partial_t B_x&=\partial_z E_y,
    &
    \partial_t E_y&=\partial_z B_x-j_y,\label{eq: irrelevant 1+1D equation 1}
    \\
    \partial_t B_z&=0,
    &
    \partial_t E_z&=-j_z .\label{eq: irrelevant 1+1D equation 2}
\end{align}
If initially
\begin{equation}
    \label{eq: irrelevant fields and currents 1+1D}
    j_y=j_z=E_y=E_z=B_x=0,\qquad B_z=B_0,
\end{equation}
then they remain at these values at all times since this field configuration satisfies Eqs.~\eqref{eq: irrelevant 1+1D equation 1} and \eqref{eq: irrelevant 1+1D equation 2}; thus, Eq.~\eqref{eq: relevant 1+1D equation} becomes the only relevant part of Maxwell's equations.   

\subsubsection{Solving FFE and NFE regions}
We now solve the FFE and NFE regions within this \(1+1\)D sector satisfying Eq.~\eqref{eq: irrelevant fields and currents 1+1D}. In this case, the fields take the form
\begin{equation}
    \label{eq: fields in the collision}
    \vec E=(E_x(t,z),0,0),\qquad
    \vec B=(0,B_y(t,z),B_0),
\end{equation}
where \(B_0\) is the constant guide field. In terms of the field tensor \(F\), this is
\begin{align}
    \label{eq: 1+1D field tensor F}
    F=E_{x}dx\wedge dt+B_{y}dz\wedge dx+B_{0} dx\wedge dy.
\end{align}
For this ansatz \(\vec{E}\cdot \vec{B}=0\) is satisfied automatically, and Gauss's law $\nabla \cdot \vec{E}=\rho$ gives $\rho=0$. 
The counter-polarized collision can be modeled by imposing reflection symmetry about the $z=0$ plane. 
Under this symmetry, the transverse electric field is even under \(z\to -z\), while the transverse magnetic field is odd:
\begin{align}
    \label{eq: EB reflection symmetry}
    E_{x}(t,z)&=E_{x}(t,-z),
    \\
    B_{y}(t,z)&=-B_{y}(t,-z).
\end{align} 
This encodes the desired counter-polarization: the two incoming waves colliding at $z=0$ have electric fields pointing in the same direction, while their transverse magnetic fields point in opposite directions. 
To make the notation simpler, we take 
\begin{align}
    B_{0}=1.
\end{align}
This can be done without loss of generality; we are simply normalizing $E_{x}$ and $B_{y}$ by the background magnetic field. To further simplify the notation, in the FFE region we write
\begin{align}
    E_{x}=E, \qquad B_{y}=B.
\end{align}
When the fields are given by Eq.~\eqref{eq: fields in the collision}, Ohm's law for FFE \eqref{eq: FFE Ohm's law} implies \(\vec j_{\rm FFE}=0\). Then, the FFE system reduces to the vacuum Maxwell equations:
\begin{align}
    \label{eq: 1+1D Maxwell 1}
    \partial_{t}B+\partial_z E&=0,
    \\
    \label{eq: 1+1D Maxwell 2}
    \partial_{t}E+\partial_z B&=0.
\end{align}
The general solution to these equations is a superposition of left- and right-moving waves:
\begin{align}
    \label{eq: FFE E field}
    E&=R(u)+L(v),
    \\
    \label{eq: FFE B field}
    B&=R(u)-L(v),
\end{align}
with $u=t-z$ and $v=t+z$.
Magnetic dominance in the FFE region requires
\begin{equation}
    \label{eq: P-FFE}
    P:=\vec{B}^2-\vec{E}^2=1-4R(u)L(v)>0.
\end{equation}
This gives the FFE part of the reduced system. 

Now let us turn to the NFE region. In the NFE region, we denote the fields by
\begin{align}
    E_{x}=\mathcal{E}, \qquad B_{y}=\mathcal{B}.
\end{align}
These fields satisfy the null condition:
\begin{equation}
    1+\mathcal{B}^2-\mathcal{E}^2=0.       \label{eq:null-condition-general}
\end{equation}
In flat spacetime, the PND $\ell$ is given by Eq.~\eqref{eq: PND flat spacetime}, and its spatial part is the Poynting direction \eqref{eq: PND flat spacetime spatial}. In the collision problem, this becomes
\begin{equation}
    \label{eq: 1+1D PND spatial}
    \vec \ell=\frac{\vec E\times \vec B}{\vec{B}^2}
    =\frac{1}{1+\mathcal{B}^2}
      \begin{pmatrix}
      0\\ -\mathcal{E}\\ \mathcal{E}\mathcal{B}
      \end{pmatrix}
    =\begin{pmatrix}
      0\\ -1/\mathcal{E}\\ \mathcal{B}/\mathcal{E}
      \end{pmatrix},
\end{equation}
where in the last step we used the null condition (\ref{eq:null-condition-general}).  
Equivalently, the full PND is
\begin{equation}
    \label{eq: 1+1D PND}
    \ell=\partial_t-\frac{1}{\mathcal{E}}\partial_y
    +\frac{\mathcal{B}}{\mathcal{E}}\partial_z .
\end{equation}
The key property of NFE is that $\ell$ is tangent to a null geodesic \eqref{eq: affine null geodesics}. 
In flat spacetime, the geodesic equation with affine parameter $\lambda$ gives
\begin{equation}
    \label{eq: 1+1D null geodesic equation}
    \partial_t\vec\ell+(\vec\ell\cdot\vec\nabla)\vec\ell=\frac{d\vec{\ell}}{d\lambda}=0.
\end{equation}
Then, from the $y$ and $z$ components of this equation, we find
\begin{align}
    \label{eq: 1+1D E B constant along ell}
    \frac{d\mathcal{E}}{d\lambda}=0, \quad \frac{d\mathcal{B}}{d\lambda}=0
\end{align}
Thus, in the planar NFE region, both \(\mathcal{E}\) and \(\mathcal{B}\) are constant along the null geodesics. For this reduced \(1+1\)-dimensional setup, we do not need the full construction discussed in Sec.~\ref{sec: initial value formulation}.

In the NFE region, the current $j_{x}$ need not be zero. In fact, using Eq.~\eqref{eq: relevant 1+1D equation}, one may write 
\begin{align}
    j_{x}=-\partial_{z}\mathcal{B}-\partial_{t}\mathcal{E}.
\end{align}

\subsubsection{Triangular wave collision and birth of NFE region}
Due to the reflection symmetry, it is sufficient to work in the $z>0$ region. The solution for the $z<0$ region is obtained from the $z>0$ solution by applying Eq.~\eqref{eq: EB reflection symmetry}. 
Let $f(s)$ denote the incoming wave packet before the collision and choose $t=0$ to be the time when the two waves start to collide.  In the $z>0$ region, the fields before first contact are then given by
\begin{align}
    E&=f(v),
    \\
    B&=-f(v).
\end{align}
We take $f(s)$ to be a triangular wave packet of height $A$ and width $W$:
\begin{equation}
    f(s)=
    \begin{cases}
        \dfrac{2A}{W}s, & 0\le s\le W/2, \quad\text{(rising branch)}\\[6pt]
        \dfrac{2A}{W}(W-s), & W/2\le s\le W, \text{(falling branch)}\\[6pt]
        0, & \text{otherwise}.
    \end{cases}                                      \label{eq:triangle-f}
\end{equation}
It is convenient to define
\begin{equation}
    \tauo:=\frac{W}{4A}.                      \label{eq:tau0-def}
\end{equation}
For $A<1/2$, the full collision process is governed by Eqs.~(\ref{eq: 1+1D Maxwell 1}) and (\ref{eq: 1+1D Maxwell 2}) since the magnetic dominance condition (\ref{eq: P-FFE}) is satisfied throughout the entire collision. In other words, this collision process is identical to the vacuum evolution, where two waves are simply superposed.
For $A>1/2$, by contrast, the vacuum evolution must be replaced by NFE once a $P=0$ region forms. 
Before this happens, the fields are given by 
\begin{equation}
    \label{eq:early-vacuum-wave}
    E=f(u)+f(v),\qquad B=f(u)-f(v).
\end{equation}
As shown in Appendix~\ref{app: vacuum collision}, when $A>1/2$, the $P=0$ surface first appears as a spacelike surface $\Birth$ given by Eq.~(\ref{eq:boundary-B appendix}).
The expression (\ref{eq:boundary-B appendix}) for $\Birth$ can be conveniently rewritten using a parameter $\sigma$ defined by $\sigma=v$. With this parameter choice, the birth front $\Birth$ is given by 
\begin{equation}
    \Birth=
    \left\{
    u=\frac{\tau_{0}^{2}}{\sigma},\quad
    v=\sigma
    \;\middle|\;
    \sigma\in [\tau_{0},W/2]
    \right\}.                                         \label{eq:boundary-B}
\end{equation}
We use the subscript $\birth$ for the coordinates on $\Birth$. Then,
\begin{equation}
    u_{\birth}(\sigma)=\frac{\tauo^2}{\sigma},
    \qquad
    v_{\birth}(\sigma)=\sigma,
\end{equation}
and in terms of \(t=(u+v)/2\) and \(z=(v-u)/2\), on $\Birth$,
\begin{equation}
    t_{\birth}(\sigma)=\frac12\left(\sigma+\frac{\tauo^2}{\sigma}\right),
    \qquad
    z_{\birth}(\sigma)=\frac12\left(\sigma-\frac{\tauo^2}{\sigma}\right).       \label{eq:birth-point}
\end{equation}

The field data on \(\Birth\) are prescribed from the FFE solution (\ref{eq:early-vacuum-wave}). 
On $\Birth$, the incoming and outgoing waves are both on the rising branch, and hence the field data are given by
\begin{align}
    \mathcal{E}(\sigma)
    &=f(u_{\birth})+f(v_{\birth})
      =\frac{\tauo^2+\sigma^2}{2\tauo\sigma},                 \label{eq:EBirth}\\
   \mathcal{B}(\sigma)
    &=f(u_{\birth})-f(v_{\birth})
      =\frac{\tauo^2-\sigma^2}{2\tauo\sigma}.                 \label{eq:BBirth}
\end{align}
Due to Eqs.~\eqref{eq: 1+1D E B constant along ell}, \(\mathcal{E}\) and \(\mathcal{B}\) are constant along each null geodesic labeled by $\sigma$ launched from $\Birth$ in the direction of $\ell$. The PND on $\Birth$ is given by 
\begin{equation}
    \ell_{\birth}(\sigma)
    =\partial_t-\frac{1}{\calE(\sigma)}\partial_y
    +U(\sigma)\partial_z .
\end{equation}
where $U(\sigma)$ is given by 
\begin{equation}
    U(\sigma)=\frac{\calB(\sigma)}{\calE(\sigma)}
    =\frac{\tauo^2-\sigma^2}{\tauo^2+\sigma^2}.               \label{eq:U-sigma}
\end{equation}
In flat spacetime, geodesics are straight lines; thus, choosing an affine parameter $\lambda$ such that the rays meet $\Birth$ at $\lambda=0$, the geodesic congruence is given by
\begin{align}
    t(\lambda;\sigma)&=t_{\birth}(\sigma)+\lambda,\label{eq:ray-t-lambda}
    \\
    z(\lambda;\sigma)&=z_{\birth}(\sigma)+U(\sigma)\lambda,       \label{eq:ray-z-lambda}
    \\
    y(\lambda;\sigma)&=y_0-\frac{\lambda}{\calE(\sigma)}.
    \label{eq:ray-y-lambda}
\end{align}
Due to planar symmetry, the \(y\)-part of the congruence is irrelevant for this problem.

Thus, to determine the NFE fields at a spacetime point $(t,z)$, one first uses Eqs.~\eqref{eq:ray-t-lambda} and \eqref{eq:ray-z-lambda} to find the label $\sigma$ of the null geodesic passing through $(t,z)$. Then, the fields are obtained by transporting birth front data (\ref{eq:EBirth}) and (\ref{eq:BBirth}) along that ray.

\subsubsection{Plasma energy in the NFE region}
Here, we derive the stress tensor of the plasma in the NFE region using the expressions for $\varphi_{\rm NFE}$ and $\varphi_{\rm m}$ defined in Eqs.~\eqref{eq: T_matter null dust} and \eqref{eq: NFE stress energy tensor}.
With our choice of affine parameter, $\ell$ takes the form given in Eq.~\eqref{eq: 1+1D PND} so that $\ell_{a}(\partial_{t})^{a}=-1$ is satisfied.
Then, $\varphi_{\rm NFE}$ can be interpreted as the energy density $\epsilon$ in the NFE region measured by an inertial observer $\partial_{t}$ since using Eq.~\eqref{eq: T_matter null dust} and $\ell_{a}(\partial_{t})^{a}=-1$ leads to 
\begin{align}
    \epsilon=T^{\rm NFE}_{ab}(\partial_{t})^{a}(\partial_{t})^{b}=\varphi_{\rm NFE}. 
\end{align}
Both EM fields and matter take the form of the null dust \eqref{eq: T_em null dust} and \eqref{eq: T_matter null dust}; therefore, the same interpretation applies to $\varphi_{\rm EM}$ and $\varphi_{\rm m}$. They are the energy density of the EM fields and matter, respectively. 
The EM field energy density in the NFE region can be calculated directly using Eq.~\eqref{eq: T_EM}. 
Using Eq.~\eqref{eq:null-condition-general} which implies $\vec{E}^{2}=\vec{B}^{2}$, we have
\begin{equation}
    \phiem=T_{tt}^{\rm NFE}=\frac12(\vec E^2+\vec B^2)=\mathcal{E}^2.
\end{equation}
For the triangular collision, the null geodesic congruence is given by Eqs.~\eqref{eq:ray-t-lambda}--\eqref{eq:ray-y-lambda}. The expansion $\theta$ is then calculated to be
\begin{align}
    \theta=\nabla_a\ell^a
    &=\partial_z\ell^z \notag
    =\partial_z U(\sigma),
\end{align}
where the $z$ derivative is taken with fixed \(t\); thus, the chain rule gives
\begin{align}
    \partial_z U(\sigma)
    &=
    U'(\sigma)
    \left(\frac{\partial\sigma}{\partial z}\right)_t . 
\end{align}
To compute \((\partial\sigma/\partial z)_t\), first eliminate the affine parameter $\lambda$ from Eqs.~\eqref{eq:ray-t-lambda} and \eqref{eq:ray-z-lambda} and write \(z\) in terms of
\(t\) and \(\sigma\):
\begin{align}
    z(t,\sigma)
    &=
    z_{\birth}(\sigma)
    +U(\sigma)\left(t-t_{\birth}(\sigma)\right).
\end{align}
Then
\begin{align}
    \left(\frac{\partial\sigma}{\partial z}\right)_t
    =\frac{1}{\left(\frac{\partial z}{\partial\sigma}\right)_t}
    =\frac{1}{z_{\birth}'(\sigma)
    +U'(\sigma)\lambda
    -U(\sigma)t_{\birth}'(\sigma)}.
\end{align}
Thus $\theta$ is
\begin{equation}
    \theta
    =\frac{U'(\sigma)}
    {z_{\birth}'(\sigma)-U(\sigma)t_{\birth}'(\sigma)
    +U'(\sigma)\lambda}.
    \label{eq:ell-expansion}
\end{equation}
Substituting this expression into Eq.~\eqref{eq:phi-general-integral} gives
\begin{align}
    \ln\frac{\varphi_{\rm NFE}(\lambda;\sigma)}{\varphi_{\rm NFE}(0;\sigma)}
    &=
    -\int_0^\lambda
    \frac{U'(\sigma)\,\dd\bar\lambda}
    {z_{\birth}'(\sigma)-U(\sigma)t_{\birth}'(\sigma)
    +U'(\sigma)\bar\lambda} \notag\\
    &=
    -\ln
    \frac{
    z_{\birth}'(\sigma)-U(\sigma)t_{\birth}'(\sigma)
    +U'(\sigma)\lambda}
    {z_{\birth}'(\sigma)-U(\sigma)t_{\birth}'(\sigma)} .
\end{align}
At $\lambda=0$, the congruence is on the birth front $\Birth$ where the matter energy is zero, so
\(\varphi_{\rm NFE}(0;\sigma)=\phiem(0;\sigma)=\calE^2(\sigma)\).  Thus
\begin{equation}
    \varphi_{\rm NFE}(\lambda;\sigma)
    =\calE^2(\sigma)
    \frac{
    z_{\birth}'(\sigma)-U(\sigma)t_{\birth}'(\sigma)}
    {z_{\birth}'(\sigma)-U(\sigma)t_{\birth}'(\sigma)
    +U'(\sigma)\lambda}.
                                                               \label{eq:phi-lambda}
\end{equation}
Using the explicit birth-front solution in Eqs.~\eqref{eq:birth-point}-- \eqref{eq:BBirth}, and \eqref{eq:U-sigma}, the plasma energy density $\epsilon_{\rm m}$ becomes
\begin{align}
   \epsilon_{\rm m}= \phim(\lambda;\sigma)
    &=\varphi_{\rm  NFE}(\lambda;\sigma)-\calE^2(\sigma) \notag\\
    &=
    \frac{-U'(\sigma)\calE^2(\sigma)\lambda}
    {z_{\birth}'(\sigma)-U(\sigma)t_{\birth}'(\sigma)
    +U'(\sigma)\lambda} \notag\\
    &=
    \frac{\lambda\sigma(\tauo^2+\sigma^2)^2}
    {(\sigma^4+\tauo^4)(\tauo^2+\sigma^2)
    -4\tauo^2\sigma^3\lambda}
    .
       \label{eq:phim-lambda}
\end{align}
The full stress tensor of the plasma follows from Eq.~\eqref{eq: T_matter null dust}.

\subsubsection{Interface law and outer interface}
Once the NFE region is born from $\Birth$, the FFE/NFE interface evolves subject to the pullback continuity \eqref{eq: interface pullback continuity general} and the interface kinetic law \eqref{eq: surface stress S eq1}--\eqref{eq: surface stress S eq2}. Since the birth front $\Birth$ ends at $\sigma=W/2$, the first such interface emerges from this location. 
We refer to the interface emerging from $\sigma=W/2$ as the ``outer interface'' (since the ``inner interface'' also emerges from $z=0$), and denote it by $I_{+}$. 
By planar symmetry, the outer interface is also a plane and can be written as $z=z_{+}(t)$ with velocity $V_{+}=dz_{+}/dt$. 
To parametrize the outer interface, let $\lambda_{+}(\sigma)$ be the value of the affine parameter such that the ray with $\sigma$ meets the interface. Then, the location of the interface is written as
\begin{align}
    \label{eq: t interface}
    t_+(\sigma)&=t_{\birth}(\sigma)+\lambda_+(\sigma),
    \\
    \label{eq: z interface}
    z_+(\sigma)&=z_{\birth}(\sigma)+U(\sigma)\lambda_+(\sigma) ,
\end{align}
and also
\begin{align}
    \label{eq: u interface}
    u_{+}(\sigma)&=\frac{\tau_{0}^{2}}{\sigma}+(1-U)\lambda_{+}(\sigma),
    \\
    \label{eq: v interface}
    v_{+}(\sigma)&=\sigma+(1+U)\lambda_{+}(\sigma).
\end{align}
The corresponding interface normal one-form \(m_a\) is written as
\begin{equation}
    \label{eq: interface normal}
    m=\frac{\dd z-V_+\dd t}{\sqrt{1-V_+^{2}}}.
\end{equation} 
The pullback $\mathcal{F}_{I_{+}}$ of $F$ to the outer interface is given by 
\begin{align}
    \mathcal{F}_{I_{+}}&=(E_{x}-V_{+}B_{y})dx\wedge dt +dx\wedge dy
\end{align}
where $(t,x,y)$ are chosen to be the intrinsic coordinates of the interface.  
Therefore, the continuity of $\mathcal{F}_{I_{+}}$ across the outer interface gives
\begin{align}
    \label{eq: interface pullback continuity}
    E-V_+ B
    &=
    \calE-V_+\calB.
\end{align}
As shown in Sec.~\ref{subsec:junction condition}, the interface law requires
the surface-stress tensor \(S^{AB}\), which in general requires a model of the
plasma in this layer.
Instead of modeling \(S^{AB}\), we use
Eq.~\eqref{eq: surface stress S eq2} to motivate a simple ad hoc closure.  For
the planar interface \(z=z_+(t)\), define
\(\gamma_+=(1-V_+^2)^{-1/2}\).  With intrinsic coordinates \((t,x,y)\), the
tangent vector along the interface motion is
\begin{equation}
    e_t{}^a=(\partial_t)^a+V_+(\partial_z)^a ,
\end{equation}
and the unit normal one-form is Eq.~\eqref{eq: interface normal}.  A direct
calculation gives
\begin{equation}
    K_{tt}=-\gamma_+\dot V_+,
    \qquad
    K_{tx}=K_{ty}=K_{xx}=K_{xy}=K_{yy}=0 .
\end{equation}
Thus the curvature-force term \(K_{AB}S^{AB}\) in
Eq.~\eqref{eq: surface stress S eq2} is controlled by the acceleration of the
interface.  Here, we approximate the interface locally as
moving with constant velocity, so that \(\dot V_+\simeq0\) and hence
\(K_{AB}S^{AB}\simeq0\).  Under this approximation,
Eq.~\eqref{eq: surface stress S eq2} reduces to the continuity of normal
pressure,
\begin{align}
    \label{eq: continuous normal pressure}
    m_{a}m_{b}[T^{ab}]&=0. 
\end{align}
This is not meant to be understood as an exact interface law, since the
interface velocity obtained below is not strictly constant.  Rather, it is a condition that is weak enough compared to Eq.~\eqref{eq:full-stress-matching-interface} so that the discontinuity of the field across the interface is allowed but still requires no additional parameter. Using Eq.~\eqref{eq: interface normal}, the continuous normal
pressure condition \eqref{eq: continuous normal pressure} becomes
\begin{align}
    0&
      =[T^{zz}-2V_{+}T^{tz}+V_{+}^{2}T^{tt}]. \label{eq:normal-pressure}
\end{align}
On the FFE side, the stress-energy tensor is purely electromagnetic with relevant components given by
\begin{align}
    T_{\rm FFE}^{tt}&=\frac12\left(E^2+B^2+1\right) \label{eq: Ttt FFE},
    \\
    T_{\rm FFE}^{tz}&=EB \label{eq: Ttz FFE},
    \\
    T_{\rm FFE}^{zz}&=\frac12\left(E^2+B^2-1\right) \label{eq: Tzz FFE}.
\end{align}
On the NFE side, the total stress tensor has the null dust form~\eqref{eq: T_matter null dust} with $\ell$ given by Eqs.~\eqref{eq: 1+1D PND} and \eqref{eq: 1+1D PND spatial}, so the relevant components of $T^{ab}_{\rm NFE}$ are
\begin{align}
     T_{\rm NFE}^{tt}&=(\mathcal{E}^2+\phim),
    \\
    T_{\rm NFE}^{tz}&=(\mathcal{E}^2+\phim) U\label{eq: Ttz NFE},
    \\
    T_{\rm NFE}^{zz}&=(\mathcal{E}^2+\phim) U^{2} \label{eq: Tzz NFE},
\end{align}
where $\mathcal{E}$, $U$, and $\phim$ are given by Eqs.~\eqref{eq:EBirth}, \eqref{eq:U-sigma}, and \eqref{eq:phim-lambda}, respectively. 
On the NFE side, the fields are obtained by the transport of the data on $\Birth$ along Eqs.~\eqref{eq:ray-t-lambda}--\eqref{eq:ray-y-lambda}. On the FFE side of the interface, the fields are given by Eq.~\eqref{eq: FFE E field} and \eqref{eq: FFE B field}, which are a superposition of a left-moving wave $L(v)$ and a right-moving wave $R(u)$. 
This right and left split in the FFE region gives a clear local picture of the effect of the interface conditions. 
On the FFE side of the outer interface, the left-moving wave $L(v)$ is already prescribed by the incoming wave and the fields on the NFE side are also given. Thus, the only unknown is the right-moving wave $R(u)$, which is determined at the interface to satisfy the conditions \eqref{eq: interface pullback continuity} and \eqref{eq:normal-pressure}. 
Substituting Eqs.~\eqref{eq: FFE E field} and \eqref{eq: FFE B field} into Eq.~\eqref{eq: interface pullback continuity} and solving for $R(u_{+})$ yields
\begin{align}
    \label{eq: R outer interface}
    R(u_{+})&=\frac{\calE(1-V_{+}U)-L(v_{+})(1+V_{+})}{1-V_{+}}.
\end{align}
It is convenient to define the magnetic jump by
\begin{equation}
    \Delta:=B-\calB.
\end{equation}
Then, using Eq.~\eqref{eq: interface pullback continuity}, we can write the FFE fields in terms of the NFE fields, the interface velocity $V_{+}$, and the magnetic jump $\Delta$ as 
\begin{equation}
    E=\calE+V_+\Delta,
    \qquad
    B=\calB+\Delta .       \label{eq:field-jump-Delta}
\end{equation}
By plugging Eqs.~\eqref{eq: Ttt FFE}--\eqref{eq: Tzz NFE} into Eq.~\eqref{eq:normal-pressure} and eliminating $E$ and $B$ using Eq.~\eqref{eq:field-jump-Delta},  we obtain the following quadratic equation for $\Delta$:
\begin{align}
    0=
    \frac{1}{2}(1-V_{+}^{2})^{2}\Delta^{2}
    +\calE(1-V_{+}^{2})(U-V_{+})\Delta
    -\phim(U-V_{+})^{2}.
    \label{eq: Delta equation}
\end{align}
The physical root is the one that behaves as \(\Delta\to0\) when \(\phim\to0\):
\begin{align}
    \Delta&=
    \frac{\calE(U-V_{+})}{1-V_{+}^{2}}
    \left(\sqrt{1+\frac{2\phim}{\calE^2}}-1\right).
    \label{eq:Delta-conservative}
\end{align}
This equation relates the magnetic field jump and the matter energy density on the NFE side. In general, $\phim$ is nonzero; thus, the continuous pressure condition \eqref{eq: continuous normal pressure} gives the discontinuity in the field across the interface. 
The magnetic jump can also be written in terms of the data from the FFE side as 
\begin{align}
    \Delta&=B-\calB\notag
    \\
    &=R(u_{+})-L(v_{+})-\calE U\notag
    \\
    &=\frac{\calE(1-U)-2L(v_{+})}{1-V_{+}},\notag
    \\
    &=\frac{G(L;\sigma)}{1-V_{+}}.
    \label{eq: magnetic jump in terms of G}
\end{align}
From the first line to the second, we have used Eqs.~\eqref{eq: FFE B field} and \eqref{eq:U-sigma}, and from the second to the third line, we have used Eq.~\eqref{eq: R outer interface}. The quantity $G(L;\sigma)$ is defined by
\begin{align}
    \label{eq: G-def}
    G_{+}(L;\sigma)=\calE(1-U)-2L(v_{+}).
\end{align}
By plugging Eq.~\eqref{eq: magnetic jump in terms of G} into Eq.~\eqref{eq:Delta-conservative}, and solving it for $V_{+}$, we find
\begin{equation}
    V_{+}(L;\sigma)
    =
    \frac{\left(\sqrt{\mathcal{E}^{2}+2\phim(\lambda_{+};\sigma)}-\mathcal{E}\right)U- G_{+}(L;\sigma)}
    {\sqrt{\mathcal{E}^{2}+2\phim(\lambda_{+};\sigma)}-\mathcal{E}+ G_{+}(L;\sigma)}.
          \label{eq:VIplus-conservative}
\end{equation}

To determine the interface, note that the velocity of the interface can be written as 
\begin{align}
        \label{eq: define Vplus}
    V_{+}=\dfrac{dz_{+}/d\sigma}{dt_{+}/d\sigma}.
\end{align}
Plugging Eqs.~\eqref{eq: t interface} and \eqref{eq: z interface} into \eqref{eq: define Vplus} and solving for \(\lambda_+'(\sigma)\) gives 
\begin{equation}
    \lambda_+'(\sigma)
    =
    \frac{z_{\birth}'(\sigma)+U'(\sigma)\lambda_+(\sigma)
    - t_{\birth}'(\sigma) V_+(L;\sigma)}
    {V_+(L;\sigma)-U(\sigma)}.
     \label{eq:lambda-plus-kinematic}
\end{equation}
The only remaining input for this equation is the data for the incoming wave $L(v)$. 
For the outer interface, $L(v_{+})$ is given by the falling branch of the incoming wave \eqref{eq:triangle-f}: 
\begin{align}
    \label{eq: L(v+) outer interface}
    L(v_{+})=\frac{W-v_{+}}{2\tau_{0}}
\end{align} 
where $v_{+}$ is given by Eq.~\eqref{eq: v interface}.
Equation~\eqref{eq: L(v+) outer interface} shows that $L(v_{+})$ only depends on $\lambda_{+}(\sigma)$ and $\sigma$. Therefore Eq.~\eqref{eq:lambda-plus-kinematic} together with Eq.~\eqref{eq: L(v+) outer interface} forms a local ODE for $\lambda_{+}(\sigma)$.  
The boundary condition for this ODE is obtained by the fact that the outer interface $I_{+}$ and the birth front $\Birth$ meet at $\sigma=W/2$; thus
\begin{align}
    \lambda_{+}\left(\frac{W}{2}\right)=0. 
\end{align}
Once this ODE is solved, the outer interface is reconstructed from Eqs.~\eqref{eq: t interface} and \eqref{eq: z interface}.

The reconstruction of the fields proceeds as follows. On the NFE side, the fields are already known. On the FFE side, the right-moving wave $R(u)$ is determined by Eq.~\eqref{eq: R outer interface}.  More specifically, first, define $\varsigma_{+}(x)$ by the inverse function of $u_{+}(\sigma)$, namely
\begin{align}
    u_{+}(\varsigma_{+}(x))=x.
\end{align}
Let $\mathcal{R}(\sigma)$ be the solution for the right-moving wave given by Eq.~\eqref{eq: R outer interface} as a function of $\sigma$. Then, the right-moving wave is given by
\begin{align}
    R(u)=\mathcal{R}(\varsigma_{+}(u)).
\end{align}
The left-moving wave $L(v)$ is the prescribed data $f(v)$; thus, the total fields are given by 
\begin{align}
     E&=\mathcal{R}(\varsigma_{+}(u))+f(v),\\
    B&=\mathcal{R}(\varsigma_{+}(u))-f(v).              \label{eq:outer-FFE-fields}
\end{align}

\subsubsection{Current sheet formation}
The NFE description is valid only before the congruence \eqref{eq:ray-t-lambda}--\eqref{eq:ray-y-lambda} develops conjugate points, after which the newly formed current sheet replaces the NFE description at the location of the conjugate point. In this reflection-symmetric collision, the rays launched from $\Birth$ move towards the $z=0$ plane since $U$ is negative; thus, the conjugate point formation occurs when the ray hits the center $z=0$. 
Setting Eq.~\eqref{eq:ray-z-lambda} to zero yields the center-hit affine parameter $\lambda_h (\sigma)$:
\begin{equation}
    z_{\birth}+U\lambda_h=0.
\end{equation}
Using Eqs.~\eqref{eq:U-sigma} and \eqref{eq: z interface}, $\lambda_{h}$ is given by
\begin{equation}
    \lambda_h(\sigma)=\frac{\tauo^2+\sigma^2}{2\sigma}.
\end{equation}
The center-hit time \(t_{h}(\sigma)\) for the ray labeled by \(\sigma\) is given by
\begin{equation}
    t_h(\sigma)=t_{\birth}(\sigma)+\lambda_h(\sigma)
    =\sigma+\frac{\tauo^2}{\sigma}.                       \label{eq:th-sigma}
\end{equation}
Since the range of $\sigma$ is $\tau_{0}\leq\sigma\leq W/2$ and $t_{h}(\sigma)$ is an increasing function of $\sigma$ in this range, the earliest conjugate point formation time is obtained at $\sigma=\tau_{0}$. Therefore the current sheet formation time $t_{\rm sheet}$ is
\begin{equation}
    t_{\rm sheet}=2\tau_{0}=\frac{W}{2A}.         \label{eq:tsheet}
\end{equation}

Precisely speaking, for a current sheet to form, at least one ray must hit the center before it exits through the outer interface. In other words, the conjugate formation requires the existence of some $\sigma$ such that 
\begin{equation}
     \lambda_h(\sigma)<\lambda_{+}(\sigma).
        \label{eq:conservative-sheet-criterion}
\end{equation}
By numerically solving Eq.~\eqref{eq:lambda-plus-kinematic} for various values of $A$, we found the approximate threshold value of $A_{c}$ above which Eq.~\eqref{eq:conservative-sheet-criterion} holds as
\begin{align}
    A_{c}\approx0.692>\frac{1}{2}.
\end{align}
The precise value of $A_{c}$ is irrelevant since our model of the interface \eqref{eq: continuous normal pressure} is not meant to exactly represent the dynamics of the interface. 
Thus, from here we assume that $A$ is sufficiently away from $A_{c}$ to avoid the case where the current sheet formation and the interface become close.

\subsubsection{After sheet formation: inner FFE region}
Once the sheet is formed at $z=0$, we impose the current sheet boundary condition \eqref{eq: null CS}. By reflection symmetry, the current sheet stays at $z=0$ and the pullback continuity \eqref{eq: pullback continuity current sheet} is automatically satisfied. The null current sheet condition \eqref{eq: null CS} then imposes
\begin{equation}
    E_x(t,z=0)=B_0=1 .  \quad t>t_{\rm  sheet}                                \label{eq:center-law}
\end{equation}
This condition supplies the data for the FFE region, resulting in the launch of FFE waves outward from $z=0$. Physically, this can be thought of as the emission of EM fields from the current sheet as the plasma particles in the sheet constantly rearrange themselves to maintain Eq.~\eqref{eq: null CS}. 
We refer to the boundary between the newly emerged FFE region from $z=0$ and the NFE region as the ``inner interface'' and denote it by $I_{-}$. 
Then, on the $z>0$ half, the post-sheet geometry can be categorized into three regions:
\begin{align}
    0<z&<z_-(t) &&: \text{inner FFE region},\\
    z_-(t)<z&<z_+(t) &&: \text{NFE region},\\
    z_+(t)<z& &&: \text{outer FFE region},
\end{align}
where $z_{-}(t)$ is the location of the inner interface. 

Using Eq.~\eqref{eq: FFE E field}, on the sheet, the condition \eqref{eq:center-law} becomes
\begin{align}
    \label{eq: 1=R+L}
    1=R(t)+L(t),\quad t>t_{\rm sheet}
\end{align}
Thus, the wave leaving from the current sheet can be thought of as  ``sheet-modulated reflection''  of the incident wave in a manner consistent with Eq.~\eqref{eq: 1=R+L}. 

The current sheet carries a surface current $j^{a}_{\rm surf}=\delta(z)K^{a}$, with $j^{a}_{\rm surf}$ given by Eq.~\eqref{eq: current sheet current vector}. Using reflection symmetry \eqref{eq: EB reflection symmetry}, $K^{a}$ is given by
\begin{align}
   K^{a}&=[F^{az}]=-2B(t,0^{+})(\partial_{x})^{a}.
   \label{eq: 1+1D surface current}
\end{align}
The sheet also exchanges energy with the surrounding FFE region. With $S=T_{tz}$ denoting the Poynting flux in the $+z$ direction, the energy flux on the two sides $z=0^{\pm}$ of the current sheet is given by
\begin{align}
    S^{\pm}=T_{tz}=B(t,0^{\pm}).
\end{align} 
The energy absorption rate $\mathcal{P}_{\rm sheet}$ is then the net flux into $z=0$:
\begin{align}
    \mathcal{P}_{\rm sheet}=-(S^{+}-S^{-})=-2B(t,0^{+}).
    \label{eq: 1+1D energy dissipation}
\end{align}
Thus, $\mathcal{P}_{\rm sheet}>0$ if $B(t,0^{+})<0$, meaning that the sheet absorbs energy from the FFE waves. When $B(t,0^{+})>0$, the sheet returns the energy to the FFE region.  

\subsubsection{Inner interface evolution}
The inner interface satisfies the same interface conditions as the outer interface. Following the same argument used to derive Eq.~\eqref{eq: R outer interface}, the right-moving wave $R(u_{-})$ is written in terms of $L(v_{-})$ as
\begin{align}
    \label{eq: R inner interface}
    R(u_{-})=\frac{\calE(1-V_{-}U)-L(v_{-})(1+V_{-})}{1-V_{-}}
\end{align}
However, in contrast to the outer interface where the form of the left-moving wave $L(v_{+})$ is already prescribed, $L(v_{-})$ is an unknown function for the inner interface and must be determined dynamically. 
To derive the equations for the inner interface, consider the current sheet \eqref{eq: 1=R+L} from which we have 
\begin{align}
    \label{eq: L=1-R}
    L(u_{-})=1-R(u_{-}).
\end{align}
Combining Eqs.~\eqref{eq: R inner interface} and \eqref{eq: L=1-R} and solving it for $L(v_{-})$ gives the relation between $L(v_{-})$ and $L(u_{-})$,
\begin{align}
    \label{eq: L(v) in terms of L(u)}
    L(v_{-})&=\frac{\calE(1-V_{-}U)-(1-V_{-})(1-L(u_{-}))}{1+V_{-}}
\end{align}
Since $v_{-}>u_{-}$ in the $z>0$ region,  Eq.~\eqref{eq: L(v) in terms of L(u)} determines the value of $L(v_{-})$ at a later argument from its earlier value at $u_{-}$. 
Thus, for the inner interface, $\lambda_{-}(\sigma)$ and $L(v)$ are two unknowns. The equation for $\lambda_{-}(\sigma)$ takes the same form as Eq.~\eqref{eq:lambda-plus-kinematic}:
\begin{align}
     \lambda_-'(\sigma)
    =
    \frac{z_{\birth}'(\sigma)+U'(\sigma)\lambda_-(\sigma)
    - t_{\birth}'(\sigma) V_-(L;\sigma)}
    {V_-(L;\sigma)-U(\sigma)},
     \label{eq:lambda-minus-kinematic}
\end{align}
where $V_{-}(L;\sigma)$ is 
\begin{align}
     V_{-}(L;\sigma)
    =
    \frac{\left(\sqrt{\mathcal{E}^{2}+2\phim(\lambda_{-};\sigma)}-\mathcal{E}\right)U- G_{-}(L;\sigma)}
    {\sqrt{\mathcal{E}^{2}+2\phim(\lambda_{-};\sigma)}-\mathcal{E}+ G_{-}(L;\sigma)},
          \label{eq:VIminus-conservative}
\end{align}
and $G_{-}(L;\sigma)$ is given by 
\begin{align}
    \label{eq: Gminus-def}
    G_{-}(L;\sigma)=\calE(1-U)-2L(v_{-}).
\end{align}
To find the inner interface, one needs to solve \eqref{eq: L(v) in terms of L(u)} and Eq.~\eqref{eq:lambda-minus-kinematic} simultaneously with appropriate initial data for $\lambda_{-}$ and $L(v)$ at the emerging point of $I_{-}$. 
Therefore, the inner interface evolution is not purely a local ODE problem; instead, the evolution of $I_{-}$ depends on the FFE data leaving $I_{-}$ at earlier times.
Such history-dependent systems are known as functional differential equations with state-dependent delay (e.g., \cite{ HartungKrisztinWaltherWu2006}). They also arise in classical electrodynamics, for instance, in the two-body problem of charged particles interacting through the retarded electromagnetic fields \cite{Driver1963,Travis1975}.

The inner interface emerges from the first conjugate point formation $t=t_{\rm sheet}, z=0$, which corresponds to 
\begin{align}
    \label{eq: inner interface lambda initial data}
    \sigma=\tau_{0},\qquad \lambda_{-}(\tau_{0})=\tau_{0},
\end{align}
which provides the initial data for $\lambda_{-}(\sigma)$. 
The remaining initial datum is $L|_{\sigma=\tau_{0}}$, which can be found by considering the limit $\sigma=\tau_{0}+\varepsilon$ with $\varepsilon\to0$. 
In this limit, the sheet condition \eqref{eq:center-law} gives $E=1$ at the emerging point. We further assume that the magnetic field $B$ is finite at $\sigma=\tau_{0}$. 
Then, from Eq.~\eqref{eq:field-jump-Delta}, we have 
\begin{align}
    V_{-}=\frac{E-\mathcal{E}}{\Delta}=O(\varepsilon^{2}).
\end{align}
Using this expression to expand Eq.~\eqref{eq:Delta-conservative} in $\varepsilon$ using Eqs.~\eqref{eq:U-sigma} and \eqref{eq:phim-lambda}, 
we find
\begin{align}
    \Delta=-1+O(\varepsilon).
\end{align}
Since $\Delta=B-\mathcal{B}$ and $\mathcal{B}=O(\epsilon)$, we find 
\begin{align}
    B=-1 +O(\varepsilon). 
\end{align}
Using this result with $E=1$, we find that in the limit of $\sigma\to \tau_{0}$
\begin{align}
    \label{eq: inner interface L initial data}
    L|_{\sigma=\tau_{0}}&=1
\end{align}
and also 
\begin{align}
    R|_{\sigma=\tau_{0}}=0,\quad V_{-}|_{\sigma=\tau_{0}}=0.
\end{align}
Equation~\eqref{eq: inner interface L initial data} gives the initial data for solving for the inner interface. 

Once the inner interface is found, the fields are reconstructed as follows. Let $\mathcal{L}(\sigma)$ denote the left-moving wave obtained from the inner interface solution.  
Let \(\varsigma_-(x)\) be the inverse function defined by
\begin{equation}
    v_-(\varsigma_-(x))=x.                              \label{eq:varsigmaminus-def}
\end{equation}
Then, the left-moving wave is given by
\begin{align}
    L(v)=\mathcal{L}(\varsigma_-(v)).
\end{align}
The right-moving wave is obtained by using Eq.~\eqref{eq: 1=R+L}, resulting in
\begin{align}
    \label{eq: R solution inner interface}
    R(u)=1-L(u)=1-\mathcal{L}(\varsigma_-(u)).
\end{align}
Therefore, in the inner FFE region, the fields are given by
\begin{align}
    E&=1-\mathcal{L}(\varsigma_-(u))+\mathcal{L}(\varsigma_-(v)),    \label{eq:inner-E}\\
    B&=1-\mathcal{L}(\varsigma_-(u))-\mathcal{L}(\varsigma_-(v)).    \label{eq:inner-B}
\end{align}
These formulas automatically satisfy Eq.~\eqref{eq:center-law} at the current sheet when \(u=v=t\).

\subsubsection{Merging of the inner and outer interfaces and subsequent evolution}
The outer interface evolves from $\sigma=W/2$ at $\Birth$ while the inner interface starts from $t=t_{\rm  sheet}$ at $z=0$. In this collision problem, they eventually meet. 
The inner and outer interfaces merge at $\sigma=\sigma_{m}$ that satisfies the following equations simultaneously:
\begin{align}
    \label{eq: define sigma_m}
    u_{-}(\sigma_{m})&=u_{+}(\sigma_{m}),\quad
    v_{-}(\sigma_{m})&=v_{+}(\sigma_{m}).
\end{align}

After the inner and outer interfaces merge, the NFE region disappears, and from the merger point, we simply continue the solution by evolving the FFE equations with the data given at $\sigma=\sigma_{m}$\footnote{This is an effective post-merger continuation: it neglects the effects of the unresolved interface layers and their collision on the macroscopic fields.}. Since the fields are discontinuous across the FFE/NFE interfaces, the merger leaves a discontinuity of the fields in the post-merger FFE region. 

The current sheet at \(z=0\) persists to impose Eq.~\eqref{eq:center-law}; thus, after the merger, the fields are given by Eqs.~\eqref{eq: FFE E field} and \eqref{eq: FFE B field} away from the sheet, and the left- and right-moving waves are related by Eq.~\eqref{eq: 1=R+L}.

The post-merger FFE region has two stages.  The first stage is the region V in Fig.~\ref{fig:sheet-region-diagram} given by
\begin{align}
    v_{m}>u>u_{m}, \quad v>v_{m}
\end{align}
where $u_{m}=u_{-}(\sigma_{m})$ and $v_{m}=v_{-}(\sigma_{m})$. In this region, the left-moving wave is supplied by the falling branch of the incoming wave packet \eqref{eq:triangle-f}, while the right-moving wave is given by Eq.~\eqref{eq: R solution inner interface}, i.e. the sheet-modulated reflection of the left-moving wave originating from the inner interface. Thus the left- and right-moving waves are
\begin{align}
    R(u)&=1-\mathcal L(\varsigma_-(u)),&
    L(v)&=f(v), \notag
\end{align}
and hence the fields are 
\begin{align}
    \label{eq: E in region V}
    E&=1-\mathcal L(\varsigma_-(u))+f(v),\\
    \label{eq: B in region V}
    B&=1-\mathcal L(\varsigma_-(u))-f(v).
\end{align}
The second stage is the region VI in Fig.~\ref{fig:sheet-region-diagram} given by
\begin{align}
    u>v_{m},\quad v>v_{m}.
\end{align}
In this region, the right-moving wave is supplied by the sheet-modulated reflection of the incoming wave \eqref{eq:triangle-f}.  Therefore, using \eqref{eq: 1=R+L}, we find
\begin{equation}
    R(u)=1-f(u),
    \qquad
    L(v)=f(v).                                       \label{eq:direct-tail-stage}
\end{equation}
and the fields are
\begin{align}
    \label{eq: E in region VI}
    E&=1-f(u)+f(v),\\
     \label{eq: B in region VI}
    B&=1-f(u)-f(v).
\end{align}
On the sheet this gives the one-sided magnetic field
\begin{equation}
    B(t,0^+)=1-2f(t).                    \label{eq:By-sheet-after}
\end{equation}
For the triangular pulse \eqref{eq:triangle-f}, $ B(t,0^+)$ continues to be negative until it crosses zero at
\begin{equation}
    t_{\rm flip}=W-\tauo.                           \label{eq:tflip}
\end{equation}
Before this time, from Eq.~\eqref{eq: 1+1D surface current} and \eqref{eq: 1+1D energy dissipation}, the surface current is flowing in the $+x$ direction and the sheet absorbs energy from the FFE waves. 
The time $t=t_{\rm flip}$ then is a transition time at which the sign of the current in the current sheet flips and the sheet switches from absorbing to emitting energy. 
In the present model, there is no reason to treat $t=t_{\rm flip}$ as a switch-off condition for the sheet. Instead,  we let the magnetic jump pass through zero and reappear with the opposite sign.

After the incoming wave is cleared, the left- and right-moving waves are
\begin{equation}
    L(t)=0,
    \qquad
    R(t)=1,
    \qquad t\ge W.
\end{equation}
Hence on the right side of the sheet the late field is
\begin{equation}
    \label{eq: collision final field}
    E=1,
    \qquad
    B=1.
\end{equation}
So the final state is the FFE region on each side plus a persistent timelike sheet.  
The sheet can persist only as long as it carries enough stored energy for the re-emission of the FFE waves.
The model, however, does not predict the precise time of sheet disappearance. The reason is that the lifetime of the sheet depends on the microscopic plasma process inside the sheet. Since the energy flowing into the sheet is redistributed to various forms of energy such as particle acceleration, radiation, and pair creation \cite{Sironi_2014,Hakobyan_2019,Cerutti_2014,Uzdensky_2013,Guo:2023ybt},  determining the fraction of injected energy that remains available for re-emission lies outside the present FFE+NFE description.

\subsubsection{Summary of field construction}\label{subsubsec:summary of construction}
\begin{figure}[t]
    \centering
    \includegraphics[width=0.95\linewidth]{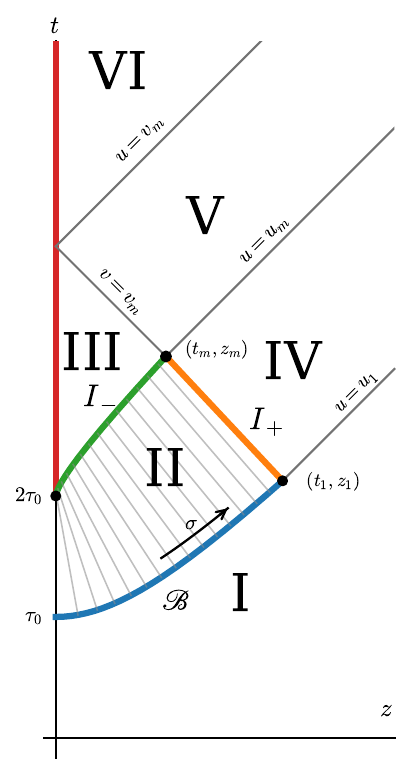}
    \caption{Spacetime structure of the collision on the
    right half-plane \(z\ge0\).  The NFE region (Region II) is born from the spacelike
    surface \(\mathscr{B}\) (blue) which is parametrized by $\sigma$. From $\Birth$, null geodesics (gray lines) are launched and the field in Region II is carried along them. The outer interface $I_{+}$ (orange) starts from the edge of $\Birth$. 
    Once the null congruence reaches \(z=0\), the current sheet (red) is formed and the inner interface $I_{-}$ (green) is launched. 
    The post-merger FFE regions V--VI are given by the FFE region and the persistent sheet. See Sec.~\ref{subsubsec:summary of construction} for more detail.}
    \label{fig:sheet-region-diagram}
\end{figure}

The construction of the regions and fields is subtle and requires lengthy calculations. Thus, it is worth summarizing the full process of FFE+NFE construction. Figure~\ref{fig:sheet-region-diagram} shows the spacetime structure of the collision of the triangular Alfv\'en waves \eqref{eq:triangle-f} on the right half-plane \(z\ge0\).  The
left half-plane is obtained by the reflection symmetry
\eqref{eq: EB reflection symmetry}. 
Region I in Fig.~\ref{fig:sheet-region-diagram} is the first FFE region before the appearance of the NFE region. In this region, the fields are given by Eq.~\eqref{eq:early-vacuum-wave}. 
As the collision process proceeds, the invariant \eqref{eq: P-FFE} reaches zero and forms the spacelike birth front $\Birth$ given in Eq.~\eqref{eq:boundary-B}, which we parametrize using $\sigma$.  
The fields on \(\Birth\) are fixed by the FFE solution
and are given by Eqs.~\eqref{eq:EBirth} and \eqref{eq:BBirth}. 
The NFE region, shown as Region II, emerges from $\Birth$ and the fields there are constructed by transporting the data on $\Birth$ along the null geodesics
\eqref{eq:ray-t-lambda}--\eqref{eq:ray-y-lambda}. 

From the outer edge of $\Birth$ corresponding to $\sigma=W/2$, the timelike outer interface $I_{+}$ starts. This interface is parametrized by
\eqref{eq: t interface} and \eqref{eq: z interface}, and its evolution is
determined by the ODE \eqref{eq:lambda-plus-kinematic}, together with the incoming-wave data \eqref{eq: L(v+) outer interface}. The FFE fields outside
\(I_+\), shown as Region IV, are reconstructed from \eqref{eq:outer-FFE-fields}.

When the transported NFE congruence reaches $z=0$, a current sheet
forms at the time given by \eqref{eq:tsheet}, which is shown by the red line at $z=0$ in Fig.~\ref{fig:sheet-region-diagram}. The sheet imposes the boundary condition \eqref{eq:center-law}, and this results in launching the FFE waves outward, creating a new FFE region shown as Region III. The FFE/NFE boundary is the timelike inner interface \(I_-\), whose location is determined by Eq.~\eqref{eq:lambda-minus-kinematic} together with the delayed relation \eqref{eq: L(v) in terms of L(u)} for $L(v)$.  Once \(I_-\) is known, the fields
in Region III are reconstructed from \eqref{eq:inner-E} and \eqref{eq:inner-B}.

The NFE region disappears when the inner and outer interfaces meet at the merger point $\sigma_{m}$ defined in Eq.~\eqref{eq: define sigma_m}.  
After the merger, the remaining evolution is purely FFE away from the sheet, while the sheet continues to impose \eqref{eq:center-law}.  In Region V, the
right-moving wave still carries the memory of the inner-interface solution, whereas the left-moving wave is the original incoming packet, giving the fields as Eqs.~\eqref{eq: E in region V} and \eqref{eq: B in region V}.  In Region VI,
this memory has cleared, and the sheet directly reflects the incoming wave. The reconstructed fields in Region VI are Eqs.~\eqref{eq: E in region VI} and \eqref{eq: B in region VI}.

\subsubsection{Comparison with the PIC simulation}
In order to test the validity of the FFE+NFE description we developed above, we conducted a one-dimensional PIC simulation using Tristan-v2 \cite{tristan_v2}.
The initial setup is two counter-propagating triangular wave packets \eqref{eq:triangle-f} on a uniform guide field.
The simulation box has
\(N_{z}=20000\) cells for the wave propagation direction and $N_{x}=N_{y}=1$ for the other directions. It is initialized with a cold neutral pair plasma with 32 particles per cell, magnetization \(\sigma=80\), and skin depth \(c/\omega_p=10\) cells. 
  No cooling was included.

\begin{figure*}[!t]
    \centering
    \includegraphics[width=0.70\textwidth]{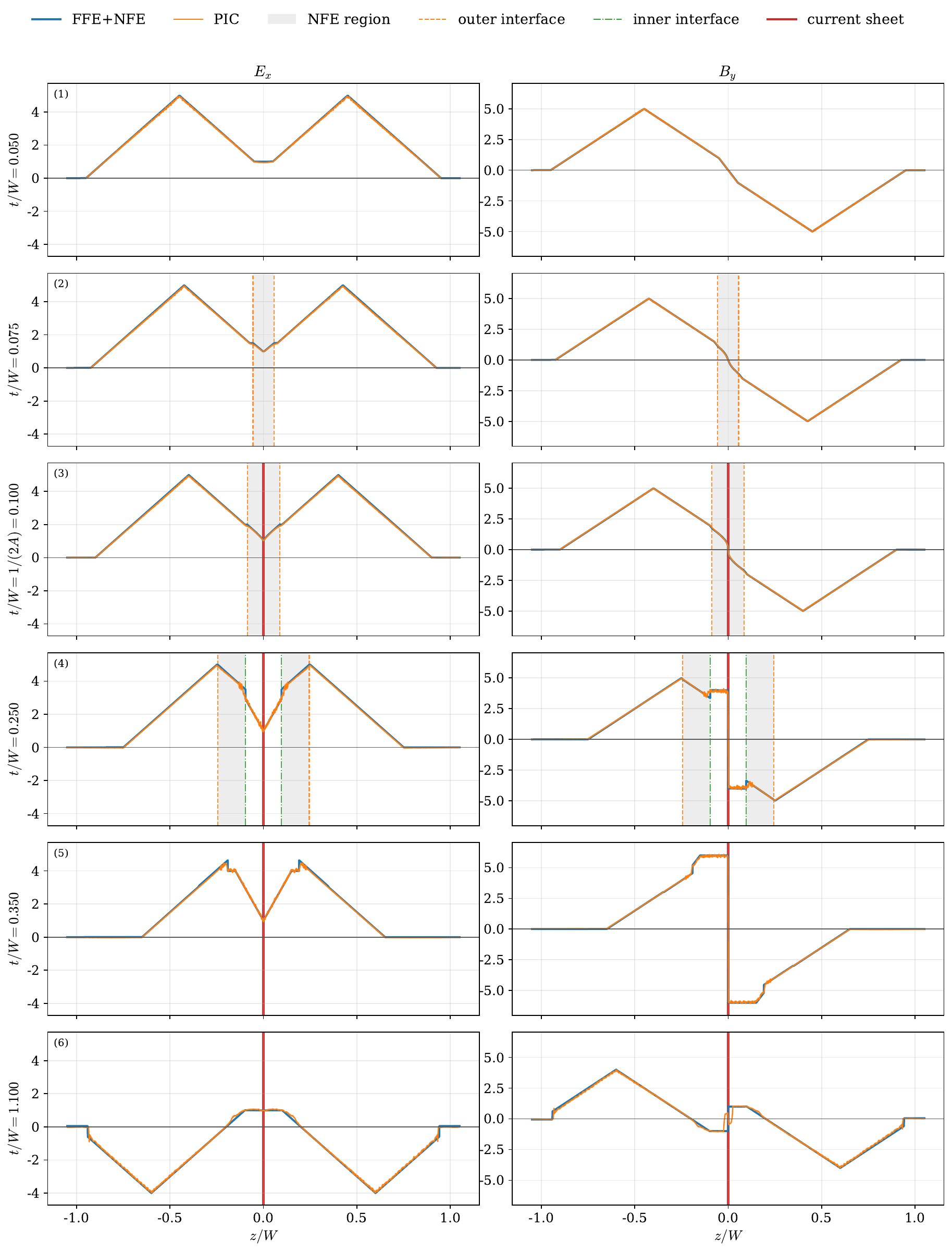}
    \caption{Comparison between the analytic model and the \(A=5\) PIC simulation. The blue line is the FFE+NFE theory and the orange line is the result of the 1+1D PIC simulation, exhibiting remarkable agreement on a macroscopic scale.  
    The gray shaded region is the NFE region, the orange and green vertical lines mark the outer and inner NFE interfaces, respectively,
    and the red line marks the current sheet.
    The full animation is available online \href{https://youtu.be/W8cHlx0qxf0}{here}.
    }
    \label{fig:pic-analytic-comparison}
\end{figure*}

Figure~\ref{fig:pic-analytic-comparison} compares the analytic \(A=5\) solution with the corresponding PIC simulation snapshots, showing agreement on the macroscopic scale.   Each row shows a representative stage of the collision.  Panel (1) is still in the vacuum-collision phase, before the NFE region has opened up.  Panel (2) shows the NFE region before a current sheet has formed.
Panel (3) shows the formation of the current sheet at time \eqref{eq:tsheet}. 
Panel (4) shows the stage in which the FFE waves are launched from the current sheet and
inner NFE interfaces move away from the origin.  Panel (5) is after the NFE region
has disappeared.  Panel (6) shows the later evolution after \(t/W=1\), where the
current sheet remains at the origin.
The gray shaded area shows the NFE region predicted by the analytic construction.
The orange vertical lines mark the outer NFE interfaces, the green vertical lines mark the inner NFE interfaces, and the red vertical line marks the current sheet and is drawn only at and after \(t=t_{\rm sheet}\).  

The close match between the FFE+NFE model and the PIC simulation shows that the macroscopic NFE description captures the essential mechanism of evolution. The PIC simulations indeed show two types of discontinuity predicted from FFE+NFE. The first one is the current sheet at $z=0$.  The other one is the field discontinuity at the FFE/NFE interface whose remnant can be most clearly seen near $z/W=0.09$ of panel (6). 
The PIC also shows the final behavior of the field given by \eqref{eq: collision final field} near $z/W \sim0$. This is because we do not include cooling or pair creation, so the energy injected in the sheet is directly returned to the field at later times.

\begin{figure*}[!t]
    \centering
    \includegraphics[width=0.96\textwidth]{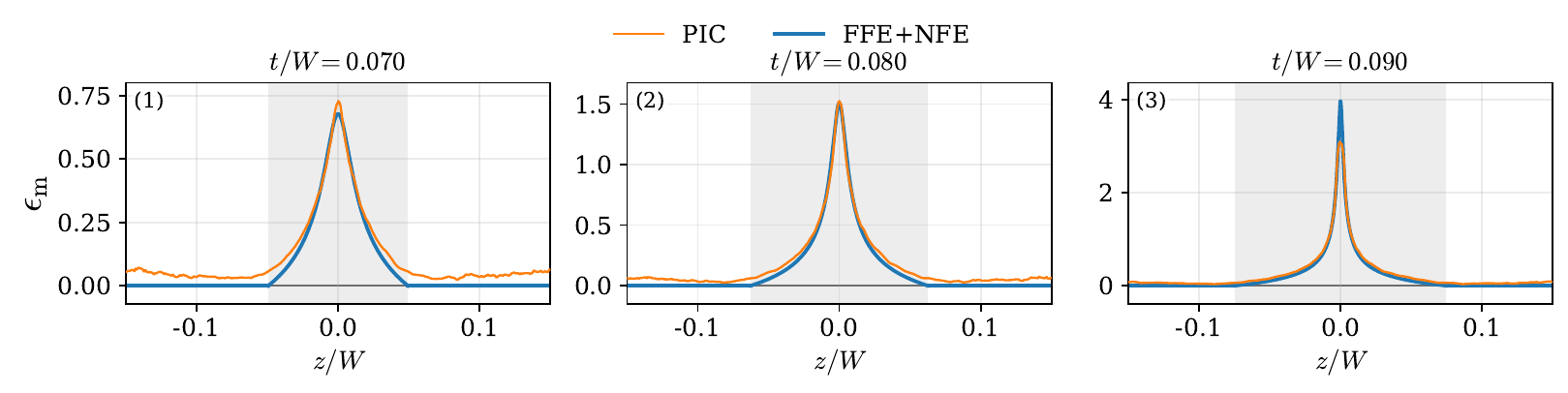}
    \caption{Comparison of the plasma energy density $\epsilon_{\rm m}$ normalized to $B^{2}_{0}$ in the \(A=5\) PIC simulation with the NFE prediction \eqref{eq:phim-lambda} before the current sheet formation. In the FFE region the matter energy density is $\epsilon_{\rm m}=0$. In the NFE region, $\epsilon_{\rm m}$ becomes larger and larger as the system becomes closer to the current sheet formation $t_{\rm sheet}/W=1/2A=0.1$. The PIC result and the NFE prediction \eqref{eq:phim-lambda} show a close match.}
    \label{fig:pic-plasma-energy-nfe}
\end{figure*}

Figure~\ref{fig:pic-plasma-energy-nfe} compares the plasma energy density measured in the PIC simulation with the NFE prediction \eqref{eq:phim-lambda}. Since we take $B_{0}=1$, the energy $\epsilon_{\rm m}$ is normalized to $B_{0}^{2}$.  Outside the NFE region, FFE predicts \(\epsilon_{\rm m}=0\) while PIC shows a finite energy density of the plasma. We can see that NFE-predicted energy \eqref{eq:phim-lambda} agree well with the PIC results, confirming that NFE can indeed capture the macroscopic plasma energy. 

Before concluding this analysis, we should mention one important caveat. In this 1+1D calculation, we do not consider the effect of tearing instability on the current sheet or the FFE/NFE interface. The higher-dimensional simulation of the collision of the Alfv\'en waves indicates that the current sheet is susceptible to tearing instability and plasmoid formation \cite{Li_Xinyu_Beloborodov_2021,N_ttil__2022,Ripperda_2021}. However, as shown in \cite{Li_Xinyu_Beloborodov_2021}, the effect of the tearing instability becomes suppressed when the amplitude of the colliding waves is small because the background magnetic field tends to stabilize the tearing mode \cite{GaleevZelenyi1976jun}. This suggests that there is a range of values of $A$ where our description remains valid. 

\subsection{NFE continuation of Adhikari's type-changing solution}
The Alfv\'en-wave collision gives an example of the FFE+NFE description including the formation of a current sheet and evolution of the FFE/NFE interface. 
Here, we study a cleaner problem of FFE-to-NFE transition using Adhikari's exact temporally type-changing solution \cite{Adhikari:2025rya}. In Minkowski coordinates \((t,x,y,z)\), it is given by 
\begin{align}
    F
    &=\sqrt{f_3+c_2}\,dt\wedge dx 
    +c_1(f_1-f_2)\,dt\wedge dz \notag\\
    &\hspace{3.5cm}
    +(f_1+f_2)\,dx\wedge dz,
    \label{eq: Adhikari FFE sol}
\end{align}
where
\begin{align}
    f_1&=e^{c_1t+x},\quad
    f_2=e^{c_1t-x},\quad
    f_3=(1-c_1^2)(f_1-f_2)^2 .
\end{align}
This solution is a genuine force-free solution with nonzero current
\begin{align}
    \label{eq: adhikari j}
    j
    &=
    -\frac{\partial_x f_3}{2\sqrt{f_3+c_2}}\,\partial_t
    +\frac{\partial_t f_3}{2\sqrt{f_3+c_2}}\,\partial_x
    \notag\\
    &\hspace{3cm}
    +(1-c_1^2)(f_2-f_1)\,\partial_z .
\end{align}
Its invariant is
\begin{align}
    F^{ab}F_{ab}=8e^{2c_1t}-2c_2 .
\end{align}
For \(-1<c_1<0\) and \(0<c_2<4\), the solution is initially magnetically dominated and becomes null on a spacelike surface
\begin{align}
    t_\birth=\frac{1}{2c_{1}}\ln\frac{c_{2}}{4}.
\end{align}
At this point, there are two natural continuations. One may continue the explicit formula \eqref{eq: Adhikari FFE sol} itself, which remains smooth across \(t=t_{\birth}\) and passes into the electrically dominated regime. In the FFE+NFE description developed here, however, we instead regard \(t=t_{\birth}\) as the birth front \(\Birth\) and use the null data there to construct the NFE evolution.

One may notice that $\Birth$ is a Cauchy surface, and thus, there is no FFE/NFE interface in the future of $\Birth$. Furthermore, as we will show shortly, there is no conjugate point formation in the NFE region. This is the reason the solution \eqref{eq: Adhikari FFE sol} is cleaner: the full continuation of the FFE solution to the entire future is possible entirely within the context of smooth NFE without introducing any phenomenological model to close the current sheet or interface laws. Also, this solution gives an instructive example of the construction discussed in Sec.~\ref{sec: initial value formulation}. 

To find the NFE continuation of Eq.~\eqref{eq: Adhikari FFE sol} after $t=t_{\birth}$, we first need to identify the initial data on the birth front. 
Let $X=x,Y=y,Z=z$ be the coordinates on $\Birth$. The fields on $\Birth$ are null and written as
\begin{align}
    F_\birth
    &=
    \sqrt{c_2}\left[
   \sqrt{1+(1-c_1^2)\sinh^2X}\,dt\wedge dX
    \right. \notag\\
    &\qquad\left.
    +c_1\sinh X\,dt\wedge dZ
    +\cosh X\,dX\wedge dZ
    \right].
\end{align}
Then, the PND on the birth front is
\begin{align}
    \label{eq: Adhikari PND in NFE}
    \ell
    &=
    \partial_t-c_1\tanh X\,\partial_x
    +\beta(X)\,\partial_z,
\end{align}
where
\begin{align}
    \label{eq: define beta}
    \beta(X)=\sqrt{1-c_1^2\tanh^2X}.
\end{align}
Using \(\lambda=t-t_\birth\) as an affine parameter, the geodesic congruence is given by
\begin{align}
    \label{eq: Adhikari ray-t}
    t&=t_\birth+\lambda,\\
    \label{eq: Adhikari ray-x}
    x&=X-c_1\lambda\tanh X,\\
    \label{eq: Adhikari ray-y}
    y&=Y\\
    \label{eq: Adhikari ray-z}
    z&=Z+\beta(X)\lambda.
\end{align}
This congruence never forms a conjugate point for $\lambda>0$. To see this, we view Eqs.~\eqref{eq: Adhikari ray-t}--\eqref{eq: Adhikari ray-z} as a ray map. 
The Jacobian of this map is given by
\begin{align}
    \label{eq: define D}
    D(\lambda,X)=\det{\frac{\partial \bm x}{\partial \bm X}}=1-c_1\lambda\,\operatorname{sech}^2X .
\end{align}
The presence of a conjugate point is indicated by the vanishing Jacobian, since $D=0$ implies the infinite focusing of the congruence. 
For $-1<c_{1}<0$, \(D>0\) holds for all \(\lambda\ge0\); therefore the NFE continuation has no conjugate point in this region. In other words, this system does not form a current sheet. 

Now, let us turn to the reconstruction of the field \(F\). 
As shown in Sec.~\ref{sec: initial value formulation}, we first need to find the pullback of $F_{\birth}$. Let \(\iota:\Birth\rightarrow M\) denote the embedding of the
birth front into spacetime. Since \(\Birth\) is the surface \(t=t_\birth\), the pullback sets \(\iota^*dt=0\). Therefore the initial two-form on \(\Birth\) is
\begin{align}
    \label{eq: Adhikari iota F_b}
    f=\iota^{*}F_\birth
    =\sqrt{c_2}\cosh X\,\dd X\wedge \dd Z .
\end{align}
The NFE fields at $(t,x,y,z)$ are obtained by writing the birth front coordinates $X,Y,Z$ in terms of $(t,x,y,z)$ connected by the geodesic \eqref{eq: Adhikari ray-t}--\eqref{eq: Adhikari ray-z}.  
Taking the differential of Eq.~\eqref{eq: Adhikari ray-x} and solving for $dX$ gives
\begin{align}
    \label{eq: Adhikari dX}
    \dd X
    =
    \frac{\dd x+c_1\tanh X\,\dd t}{D},
\end{align}
where \(\dd\lambda=\dd t\) from Eq.~\eqref{eq: Adhikari ray-t} and $D$ is given by Eq.~\eqref{eq: define D}.
A similar calculation for Eq.~\eqref{eq: Adhikari ray-z} gives
\begin{align}
    \label{eq: Adhikari dZ}
    \dd Z
    =
    \dd z-\beta(X)\,\dd t-\lambda\beta'(X)\,\dd X .
\end{align}
Therefore, by replacing $dX$ and $dZ$ in Eq.~\eqref{eq: Adhikari iota F_b} with Eqs.~\eqref{eq: Adhikari dX} and \eqref{eq: Adhikari dZ}, we find
\begin{align}
    F
    &=
    \frac{\sqrt{c_2}\cosh X}{D}
    \left[
    \beta(X)\,dt\wedge dx
    \right. \notag\\
    &\hspace{2.5cm}\left.
    +c_1\tanh X\,dt\wedge dz
    +dx\wedge dz
    \right],
    \label{eq:Adhikari-transported-F}
\end{align}
where \(X=X(t,x)\) is determined implicitly by the ray map
\eqref{eq: Adhikari ray-t} and \eqref{eq: Adhikari ray-x}. This is the
desired NFE continuation. One can also show directly that this field is null everywhere in the NFE region.

In the NFE region, matter carries a nonzero stress-energy tensor.  To find the matter energy, let us first focus on the energy of the EM fields. 
In Eq.~\eqref{eq:Adhikari-transported-F}, the electric field and magnetic field can be found by reading off the terms containing \(dt\), while the purely spatial part \(dx\wedge dz\) gives the magnetic field. Thus
\begin{align}
    \vec E^{\,2}=\vec{B}^{2}
    &=
    \frac{c_2\cosh^2X}{D^2},
\end{align}
where we have used Eq.~\eqref{eq: define beta} to simplify the expression for $\vec{E}^{2}$.
Therefore the electromagnetic energy density measured by the inertial
observer \(\partial_t\) is
\begin{align}
    T_{\rm EM}^{tt}
    =
    \frac12\left(\vec E^{\,2}+\vec B^{\,2}\right)
    =
    \frac{c_2\cosh^2X}{D^2}.
\end{align}
Since the affine parameter was chosen so that \(\ell^t=1\), $\phiem=T_{\rm EM}^{tt}$; thus, the full stress-energy tensor of the EM field is 
\begin{align}
    T_{\rm EM}^{ab}=\phiem\,\ell^a\ell^b,
    \qquad
    \phiem=\frac{c_2\cosh^2X}{D^2}.                  \label{eq:Adhikari-phiem}
\end{align}
The expansion of the congruence can be computed directly using the PND given by Eq.~\eqref{eq: Adhikari PND in NFE}. 
Since \(\ell^t=1\), \(\ell^y=0\), and \(\ell^z=\beta(X)\) has no \(z\)
dependence, only the \(x\)-derivative contributes to the expansion:
\begin{align}
    \nabla_a\ell^a
    =
    \partial_x\left[-c_1\tanh X(t,x)\right]
    =
    -c_1\operatorname{sech}^2X\,\partial_xX .
\end{align}
To evaluate \(\partial_xX\), differentiate the ray-map relation \eqref{eq: Adhikari ray-x} with respect to $x$ at fixed \(t\):
\begin{align}
    1
    =
    \left(1-c_1\lambda\,\operatorname{sech}^2X\right)\partial_xX
    =
    D\,\partial_xX .
\end{align}
Here $D$ is the Jacobian \eqref{eq: define D}.
Therefore
\begin{align}
    \nabla_a\ell^a
    =
    -\frac{c_1\operatorname{sech}^2X}{D}
    =\frac{1}{D}\frac{dD}{d\lambda}.
\end{align}
This gives
\begin{align}
    \label{eq: phiNFE divergence}
    \nabla_a(\varphi_{\rm NFE}\ell^a)
    =
    \frac{\dd\varphi_{\rm NFE}}{\dd\lambda}
    +\frac{\varphi_{\rm NFE}}{D}\frac{\dd D}{\dd\lambda}.
\end{align}
Since $\nabla_a(\varphi_{\rm NFE}\ell^a)=0$ holds from Eq.~\eqref{eq: NFE energy conservation}, Eq.~\eqref{eq: phiNFE divergence} implies
\begin{align}
    \label{eq: d phiNFE D=0}
    \frac{\dd}{\dd\lambda}\left(\varphi_{\rm NFE}D\right)=0 .
\end{align}
At the birth front $\Birth$, \(\phim(0,X)=0\).  Since \(D=1\) at
\(\lambda=0\) from Eq.~\eqref{eq: define D}, $\varphi_{\NFE}$ is equal to Eq.~\eqref{eq:Adhikari-phiem} on $\Birth$:
\begin{align}
    \varphi_{\rm NFE}(0,X)=\phiem(0,X)=c_2\cosh^2X .
\end{align}
Equation \eqref{eq: d phiNFE D=0} can then be solved for $\varphi_{\rm NFE}(\lambda,X)$, giving
\begin{align}
    \label{eq: phiNFE Adhikari}
    \varphi_{\rm NFE}(\lambda,X)=\frac{\phiem(0,X)}{D}=\frac{c_2\cosh^2X}{D}.
\end{align}
The matter contribution $\phim$ is simply the difference between $\varphi_{\rm NFE}$ and $\phiem$ as shown in Eq.~\eqref{eq: NFE matter energy}.  Using
Eqs.~\eqref{eq:Adhikari-phiem} and \eqref{eq: phiNFE Adhikari}, $\phim$ is given by 
\begin{align}
    \phim
    &=
    \varphi_{\rm NFE}-\phiem \notag\\
    &=
    \frac{c_2\cosh^2X(D-1)}{D^2}\notag\\
    &=
    -\frac{c_1c_2\lambda}{D^2}.                   \label{eq:Adhikari-phim}
\end{align}
To derive the final line, we have used Eq.~\eqref{eq: define D}. 
The energy density of the matter is then:
\begin{align}
    \label{eq:Adhikari-plasma energy}
    \epsilon_{\rm m}=\phim=-\frac{c_1c_2\lambda}{D^2}.
\end{align}
This is positive for the branch considered here because \(c_1<0,c_{2}>0\), and
\(\lambda\ge0\).

We can also calculate the Lorentz force density. Using \(\ell^a\nabla_a=d/d\lambda\), we have
\begin{align}
    \nabla_a(\phim\ell^a)
    &=
    \frac{\dd\phim}{\dd\lambda}
    +\phim\nabla_a\ell^a\notag \\
    &=
    -\frac{c_1c_2}{D^2}
    +
    \frac{c_1c_2\lambda}{D^3}\frac{\dd D}{\dd\lambda}\notag \\
    &=
    -\frac{c_1c_2}{D^3},
\end{align}
where Eq.~\eqref{eq: define D} has been used. Then, Eq.~\eqref{eq: null force matter} gives
\begin{align}
    F\indices{^a_b}j^b
    =
    -\frac{c_1c_2}{D^3}\ell^a . \label{eq:Adhikari-force-density}
\end{align}
Indeed, the Lorentz force density is proportional to the PND, as it should be in NFE. 

\subsubsection{Comparison to PIC}
We compare the FFE and its NFE continuation constructed in Eq.~\eqref{eq:Adhikari-transported-F} with a one-dimensional Tristan-v2 PIC simulation of Adhikari's type-changing solution.
The 1+1D reduction was possible since Adhikari's solution \eqref{eq: Adhikari FFE sol} depends only on $t$ and $x$ and can be treated in \(1+1\) dimensions.  
For the simulation, we chose \(c_1=-0.5\) and \(c_2=1\), for which the birth time is \(t_\birth=2\log 2\sim1.39\). The simulation has \(N_x=32000\) cells and skin depth \(c/\omega_p=10\) cells, and is initialized with magnetization \(\sigma=1000\), and a cold plasma with 16 particles per cell. 

Adhikari's solution \eqref{eq: Adhikari FFE sol} is not bounded in space, so it was necessary to truncate the solution to place it in a PIC simulation box.
In the simulation, the exact force-free solution at $t=0$ is imposed in the central region \(|x|\le2\). 
In \(|x|\ge3\), the fields are chosen to be a constant magnetically dominated configuration whose representative values are taken to be $F$ evaluated at $t=0$, $x=0$.
For \(2<|x|<3\), the fields are smoothly tapered so that they match Adhikari's solution at $x=2$ and a constant uniform field at $x=3$. We chose $t=0$ as the initialization time. At $t=0$, the electron and positron macroparticle weights were assigned locally to support the charge density given by Eq.~\eqref{eq: adhikari j}. The velocities were initialized as opposite species-dependent offsets from the drift velocity to support the required current density \eqref{eq: adhikari j}. 
After $t=0$, the system was evolved self-consistently by PIC. Even though the initialization of the particles is not motivated by an exact kinetic equilibrium, the plasma-free nature of FFE and NFE suggests that the evolution would not depend sensitively on the microscopic particle distribution function, provided that the charge and current are properly supplied. 

\begin{figure}[htbp]
    \centering
    \includegraphics[width=\columnwidth]{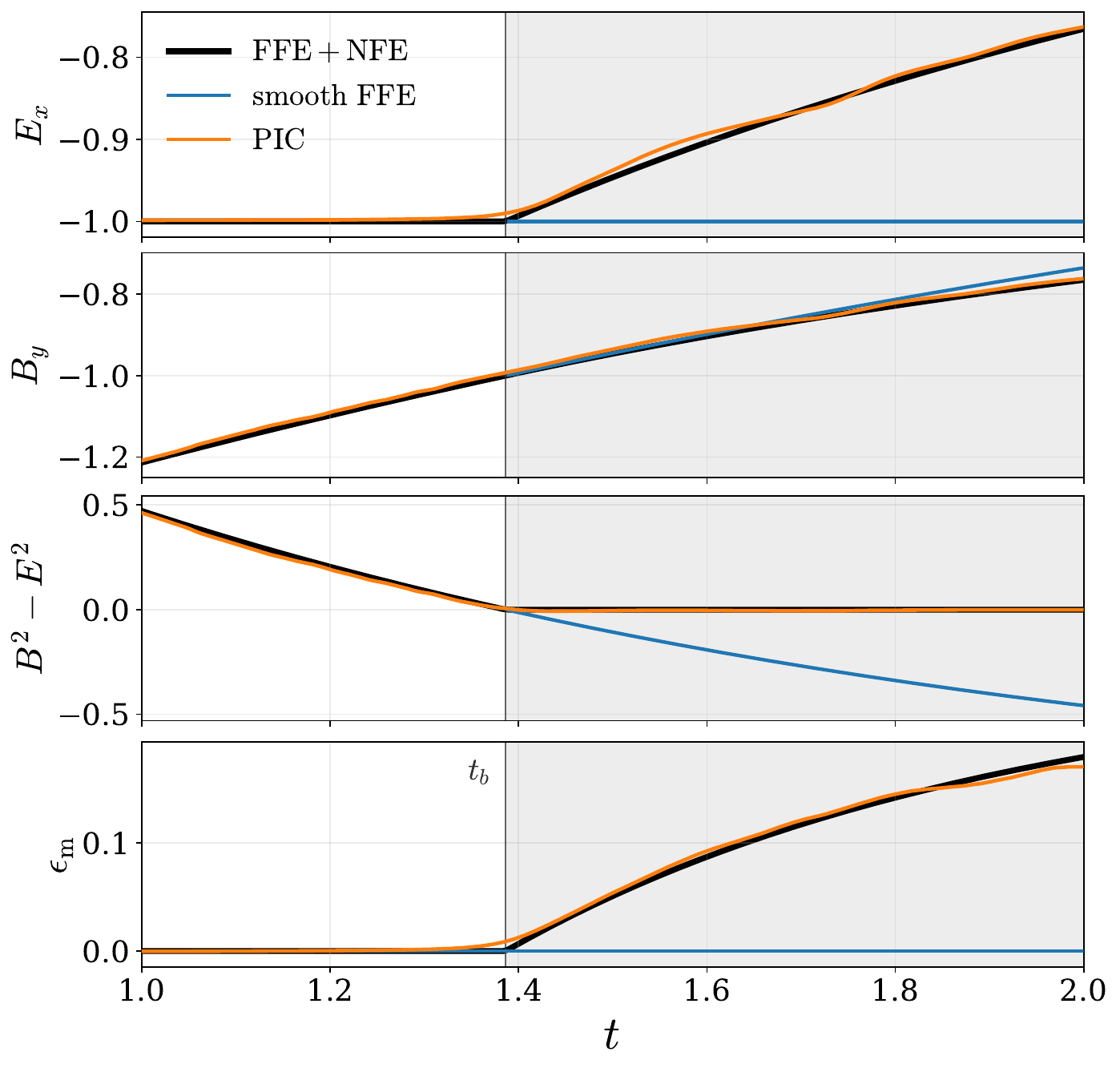}
    \caption{PIC comparison for Adhikari's type-changing solution. The NFE region is shaded in gray. The vertical line marks the birth time \(t_{\birth}\). The black curves show the FFE+NFE continuation \eqref{eq:Adhikari-transported-F}, the orange curves show the PIC result, and the blue curves show the smooth FFE continuation given by continue using Eq.~\eqref{eq: Adhikari FFE sol} in the electrically dominated region. The field panels show the nonzero components \(E_x\) and \(B_y\), together with \(\vec{B}^2-\vec{E}^2\), at \(x=0\); the other field components vanish in this reduced setup. The plasma-energy panel shows the PIC average of the plasma energy density \(\epsilon_{\rm m}\) over \(|x|\leq0.03\), compared with the NFE prediction \eqref{eq:Adhikari-plasma energy}. All panels indicate that the NFE continuation captures the evolution of the field as well as the particle energy density. }
    \label{fig:adhikari-pic-energy-summary}
\end{figure}

Figure~\ref{fig:adhikari-pic-energy-summary} shows the simulation results at the representative location \(x=0\). The first two panels show the nonzero field components \(E_x\) and \(B_y\); the other field components vanish in this reduced setup. The third panel shows the invariant $P=\vec{B}^{2}-\vec{E}^{2}$, and the fourth panel shows the plasma energy density. For the plasma energy density, in order to reduce the noise in the simulation result, we took the spatial average over the 120 cells centered at $x=0$, which corresponds to the average over \(|x|\le0.03\). The vertical line in Fig.~\ref{fig:adhikari-pic-energy-summary} marks the birth time \(t_\birth\) of the NFE region, where the smooth FFE solution reaches null, and the gray shading marks the NFE region. Before \(t_\birth\), the PIC data follow the force-free solution \eqref{eq: Adhikari FFE sol}. After \(t_\birth\), the smooth FFE continuation becomes electrically dominated, while the PIC evolution instead follows the NFE continuation \eqref{eq:Adhikari-transported-F}. 
The energy-density panel also shows that the plasma energy density predicted by NFE in Eq.~\eqref{eq:Adhikari-plasma energy} indeed captures the PIC-calculated energy density. 
All of the panels show that NFE is a better continuation of FFE than smooth FFE after the loss of magnetic dominance. 

\section{Limitations}\label{sec:limitations}
In this paper we have developed NFE and explored its various properties. 
We have found that 1) NFE is a theory that follows the conservation of the total stress-energy, with an assumption that matter takes the form of null dust, 2) it is specified by a null geodesic congruence and foliation of spacetime together with the value of the field on the field sheet, 3) an NFE solution with specified initial data on a given initial surface may be found by Lie transporting the field along the null geodesic launched from the surface in the direction of the PND, and 4) it allows the formation of field discontinuities (current sheets) within a finite time. 
Using analytical solutions to FFE that exhibit the finite-time loss of magnetic dominance, we constructed the NFE continuation of these solutions and compared them with the PIC simulations of the same initial setup.  
NFE is a parameter-free effective theory meant to describe the null electromagnetic field without resolving detailed plasma physics, and thus it has a definite domain of validity and limitations.  Here, we mention some of the limitations of the NFE continuation of FFE.

The first limitation concerns the transition from FFE to NFE. In the idealized description used here, the FFE solution reaches $F^{ab}F_{ab}=0$ on a spacelike birth front $\Birth$ and NFE is continued using data prescribed on $\Birth$ from the FFE side. 
The field on $\Birth$ is continuous across the transition, but not smooth, so its derivatives can jump. However, this does not necessarily pose any significant issues for the NFE continuation. Rather, it is a natural consequence of taking the appropriate limit 
where the plasma skin depth is unresolved and the plasma energy is completely ignored.
In a real kinetic plasma, $F^{ab}F_{ab}=0$ is never reached exactly since timelike particles can never reach null, so a finite plasma will smooth the transition over a microscopic layer. However, it can get increasingly closer to null as the electromagnetic energy scale becomes much larger than the matter energy scale. 
In this sense, the sharp FFE-to-NFE transition on $\Birth$ should be viewed as the zero-thickness limit of a more complicated but increasingly localized plasma process.

A second limitation concerns current sheets. In this paper, a current sheet is merely a boundary condition for the fields outside. This is a simple treatment reflecting the philosophy of FFE and NFE that no new parameter is introduced. The price is that the internal dynamics of the sheet is absent, and the model does not capture various kinetic processes characteristic of the sheet such as tearing or plasmoid formation \cite{Loureiro:2007gv,Uzdensky_2013,Li_Xinyu_Beloborodov_2021}, radiative cooling and pair creation \cite{Hakobyan_2019}.
For example, in the example of the Alfv\'en wave collision in Sec.~\ref{subsec: Alfven wave collision}, the formed current sheet continues to exist forever unless we turn it off by hand. In reality, however, the sheet can remain consistent with the boundary-condition description only as long as it has enough stored energy to supply the emitted FFE waves.  The fraction of injected energy that remains available for such re-emission is controlled by microscopic plasma processes; it therefore lies beyond the FFE+NFE theory. 
Relatedly, the present model does not describe the remnants left behind after a current sheet ceases to behave as an idealized boundary. Such a remnant would require additional state variables and a kinetic or fluid description of the layer. This lies outside the FFE+NFE effective theory developed here.

A third limitation concerns FFE/NFE interfaces. 
As we have shown in Sec.~\ref{subsec:junction condition}, the FFE/NFE interface is generally a kinetic layer that requires information about the surface stress-energy. 
In the example, we have treated the layer by imposing a simple ad hoc junction condition in order to avoid modeling the interface physics. Also, when two interfaces collide, as in the case when the inner and outer interfaces merge, we only treat this as a disappearance of the NFE region, and continue the solution without any consideration for the evolution of the kinetic layer. None of these treatments produced any significant deviation from the PIC simulation result, since the discontinuity of the field across the interface was small, and the simulation was done in 1+1D. However, a more concrete treatment requires an additional theory to resolve more general properties of the interface.

\section{Outlook}\label{sec:outlook}
The main consequence of this paper is that the loss of magnetic dominance in FFE need not signal the end of effective theory. FFE can indeed be continued, at least in the examples studied here, by NFE. This supports an effective theory viewpoint beyond FFE: FFE may be only one member of a larger number of effective plasma-free descriptions, each valid in a different electromagnetic regime \cite{Gralla:2018kif,Gralla:2018bvg,Jacobson:2015cia,Gruzinov:2013pva}.

One natural direction is therefore to search for effective theories beyond FFE other than NFE. The breakdown of FFE is not limited to the loss of magnetic dominance; thus,
other breakdown regimes may admit different effective descriptions. For example, near pulsar polar caps, a charge-starved region emerges, which results in a region with $\vec{E}\cdot\vec{B}\neq0$, followed by pair creation \cite{Philippov:2020jxu,Tolman:2022unu}. 
It would be interesting to explore whether there exists an effective description of such regions that is analogous in spirit to FFE and NFE: a theory in which the microscopic plasma processes are not resolved explicitly, but their leading macroscopic effect is governed by a simple principle such as a conservation law.

Another direction is the development of a better effective theory for current sheets and FFE/NFE interfaces with the minimal number of new parameters and degrees of freedom. 
In this paper, the sheet boundary condition and the interface junction condition used in the example of the Alfv\'en wave collision are chosen to be parameter-free. This is useful as it allows the full description of evolution only in terms of the field, but it cannot capture the full range of kinetic effects in the layer. A natural next step is to introduce a model of surface degrees of freedom, such as surface stress-energy, solve for their evolution, and compare the result with kinetic simulations. This would allow the theory to retain simplicity while capturing the more detailed effects not seen in the parameter-free theory. 

Astrophysical applications of FFE (e.g. black hole magnetospheres \cite{BlandfordZnajek1977may} and pulsars \cite{Gold1968may,GoldreichJulian1969aug}) also suggest the potential realization of NFE regions in these magnetospheres. Perhaps the most relevant astrophysical application of the NFE description is the monster shock \cite{Beloborodov:2022pvn}. As the initially small fast magnetosonic waves launched near the neutron star propagate outward, the strength of the wave relative to the background becomes large \cite{Yuan:2022uqt,BarrabesIsrael1991feb}. This eventually leads to the formation of a macroscopically large region with small $\vec{B}^{2}-\vec{E}^{2}$ \cite{Bernardi:2025wht,Grehan:2026vof,Vanthieghem:2024van,Yuan:2022uqt,Beloborodov:2022pvn}. This seems to be the perfect situation where the NFE description can be applied. 

It is also important to understand more systematically when the assumptions used for the geometry of the current sheet, birth front, and interface are realized dynamically.  In this paper, we have assumed that the birth front $\Birth$ is a spacelike hypersurface, and the interface $I$ and the current sheet $\mathcal{C}$ are timelike hypersurfaces. However, from generic FFE initial data, more general geometry of these objects may arise, such as tearing of the interface and sheet, a null FFE/NFE interface, and lower-dimensional versions of the current sheet.  

Finally, NFE may be useful as a tool in numerical simulations. Many force-free simulations already use prescriptions to control the loss of magnetic dominance, such as clipping the electric field \cite{Komissarov:2005xc,Spitkovsky_2006,McKinney:2006sc} or modifying Ohm's law by adding resistive or dissipative terms (e.g., \cite{Mahlmann:2020nwe,Parfrey_2012}). It would be valuable to compare such prescriptions with the NFE continuation proposed here. At least in the examples considered in this paper, NFE indeed captures the valid continuation; thus such comparison may provide a useful benchmark for assessing the numerical treatment of the loss of magnetic dominance.

\section*{Acknowledgments}
I would like to thank Sam Gralla for extensive discussions. I would also like to thank Alexander Philippov for thoughtful comments. 
This work was supported by grants from the Simons Foundation (MP-SCMPS-00001470) and the National Science Foundation (PHY-2309191, PHY-2513082).
\appendix

\section{\texorpdfstring{Ill-posedness of the \(3+1\) NFE formulation}{Ill-posedness of the 3+1 NFE formulation}}
\label{app:nfe-3plus1-illposedness}
Here we show that the \(3+1\) formulation of NFE is not well-posed.  We use the Maxwell equations
\eqref{eq: 3+1D maxwell1}--\eqref{eq: 3+1D maxwell2} and NFE Ohm's law \eqref{eq: NFE Ohm 3+1}.  Following \cite{Komissarov2002nov}, we need only focus on the local properties of the equations, so it is enough to consider a constant null background aligned with the coordinates as
\begin{equation}
    \vec E_{0}=\hat{x},\qquad
    \vec B_{0}=\hat{y},\qquad
    \vec{\ell}_{0}=\hat{z},
\end{equation}
where $\hat{x}$, $\hat{y}$, and $\hat{z}$ are unit coordinate vectors. 
Then, we perturb the field and current about the background as 
\begin{equation}
    \vec E=\vec E_{0}+\delta \vec{E} ,
    \qquad
    \vec B=\vec B_{0}+\delta \vec{B},
    \qquad
    \vec j=\delta\vec j,
\end{equation}
and keep only the first order perturbation.  The null condition \eqref{eq: 3+1 null field condition} can be linearized as
\begin{align}
    \vec E_{0}\cdot \delta\vec B + \delta\vec E\cdot\vec B_{0}=0,
    \\
    \vec B_{0}\cdot \delta\vec B-\vec E_{0}\cdot\delta\vec E=0,
\end{align}
which lead to 
\begin{align}
    \delta B_x=-\delta E_y,\qquad
    \delta B_y=\delta E_x . \label{eq: linearized constraints}
\end{align}
The linearized Maxwell equations are
\begin{equation}
    \label{eq: linearized NFE maxwell}
    \partial_t\delta\vec B=-\nabla\times\delta\vec E,
    \qquad
    \partial_t\delta\vec E=\nabla\times\delta\vec B-\delta\vec j .
\end{equation}
Linearizing NFE Ohm's law \eqref{eq: NFE Ohm 3+1} and then using Eq.~\eqref{eq: linearized constraints} gives
\begin{align}
    \delta j_x&=-\partial_x \delta E_z+\partial_y \delta B_z,
    \\
    \delta j_y&=-\partial_x \delta B_z-\partial_y \delta E_z,
    \\
    \delta j_z&=\partial_x \delta E_x+\partial_y \delta E_y+\partial_z \delta E_z .
\end{align}
Substituting these expressions into Eq.~\eqref{eq: linearized NFE maxwell}
gives the full linearized evolution system
\begin{align}
    \partial_t\delta B_x&=-\partial_y\delta E_z+\partial_z\delta E_y,+
    \\
    \partial_t\delta B_y&=-\partial_z\delta E_x+\partial_x\delta E_z,
    \\
    \partial_t\delta B_z&=-\partial_x\delta E_y+\partial_y\delta E_x,
    \\
    \partial_t\delta E_x&=\partial_x\delta E_z-\partial_z\delta B_y,
    \\
    \partial_t\delta E_y&=\partial_y\delta E_z+\partial_z\delta B_x,
    \\
    \partial_t\delta E_z&=
    \partial_x\delta B_y-\partial_y\delta B_x\notag
    \\
    &\qquad-\partial_x\delta E_x-\partial_y\delta E_y-\partial_z\delta E_z .
        \label{eq:nfe-linear-full-system}
\end{align}
Imposing the null constraints \eqref{eq: linearized constraints} removes
\(\delta B_x\) and \(\delta B_y\) as independent variables.  The same system becomes 
\begin{align}
    (\partial_t+\partial_z)\delta E_x&=\partial_x\delta E_z,
        \label{eq:nfe-linear-constrained-system1}
    \\
    (\partial_t+\partial_z)\delta E_y&=\partial_y\delta E_z,
        \label{eq:nfe-linear-constrained-system2}
    \\
    (\partial_t+\partial_z)\delta E_z&=0,
        \label{eq:nfe-linear-constrained-system3}
\end{align}
for the electric field perturbation. The perturbed magnetic field component satisfies 
\begin{align}
    \partial_t\delta B_z&=-\partial_x\delta E_y+\partial_y\delta E_x.
\end{align}
Now, we consider the Fourier ansatz with \(\vec k=(k_x,k_y,k_z)\),
\begin{align}
    \delta\vec E(t,\vec x)&=\widehat{\delta\vec E}(t)e^{i\vec k\cdot\vec x},
    \\
    \delta\vec B(t,\vec x)&=\widehat{\delta\vec B}(t)e^{i\vec k\cdot\vec x}.
\end{align}
Using Eq.~\eqref{eq:nfe-linear-constrained-system1}--\eqref{eq:nfe-linear-constrained-system3}, we obtain the equations
for the electric field perturbation:
\begin{align}
    \frac{\dd}{\dd t}
    \begin{pmatrix}
        \widehat{\delta E}_x\\
        \widehat{\delta E}_y\\
        \widehat{\delta E}_z
    \end{pmatrix}
    &=
    i
    \begin{pmatrix}
        -k_z & 0 & k_x\\
        0 & -k_z & k_y\\
        0 & 0 & -k_z
    \end{pmatrix}
    \begin{pmatrix}
        \widehat{\delta E}_x\\
        \widehat{\delta E}_y\\
        \widehat{\delta E}_z
    \end{pmatrix}.
      \label{eq:nfe-linear-fourier-matrix}
\end{align}
Solving these ODEs gives
\begin{align}
    \begin{pmatrix}
        \widehat{\delta E}_x\\
        \widehat{\delta E}_y\\
        \widehat{\delta E}_z
    \end{pmatrix}
    =
    e^{-ik_z t}
    \begin{pmatrix}
        \widehat{\delta E}_x(0)+ik_x t\,\widehat{\delta E}_z(0)\\
        \widehat{\delta E}_y(0)+ik_y t\,\widehat{\delta E}_z(0)\\
        \widehat{\delta E}_z(0)
    \end{pmatrix}.
        \label{eq:nfe-linear-fourier-sol}
\end{align}
Thus, perturbations with larger transverse wave vector grow faster, indicating
that the system is not well posed.

\section{Surface balance law for a timelike interface}
\label{app:surface-balance-law}
Here we derive Eqs.~\eqref{eq: surface stress S eq1} and
\eqref{eq: surface stress S eq2}.  
Recall that the interface \(I\) is a timelike interface, and 
\(\Phi\) is defined such that the interface $I$ is given by \(\Phi=0\) and $\Phi$ smoothly increases from NFE to the FFE side, and is further chosen such that the normal vector \(m^a=\nabla^{a}\Phi\) to the interface has unit length and points from the NFE side to the FFE side.
On the interface, the jump is given by
\begin{equation}
    [T^{ab}]=T_{\rm FFE}^{ab}-T_{\rm NFE}^{ab}.
\end{equation}
Let \(y^A\) be coordinates intrinsic to \(I\), with coordinate basis
\(e_A{}^a=\partial x^a/\partial y^A\) intrinsic to $I$.  The induced metric and extrinsic
curvature are
\begin{equation}
    h_{AB}=g_{ab}e_A{}^a e_B{}^b,
    \qquad
    K_{AB}=e_A{}^a e_B{}^b\nabla_a m_b .
\end{equation}
We include an intrinsic surface stress tensor \(S^{AB}\) on the interface and
write the total stress tensor as
\begin{align}
    T^{ab}
    =
    T^{ab}_{\rm bulk}
    +s^{ab}\delta (\Phi),
\end{align}
where
\begin{align}
    \label{eq: bulk stress tensor}
    T_{\rm bulk}^{ab}&=
    T_{\rm FFE}^{ab}\Theta(\Phi)
    +T_{\rm NFE}^{ab}\Theta(-\Phi),
    \\
    \label{eq: surface stress tensor}
     s^{ab}&=S^{AB}e_A{}^a e_B{}^b .
\end{align}
Then, the tensor \(s^{ab}\) annihilates $m^{a}$ in both indices:
\begin{equation}
    m_a s^{ab}=0,\qquad m_b s^{ab}=0 .
\end{equation}
The bulk stresses are conserved away from the interface since both FFE and NFE satisfy $\nabla_{a}T^{ab}_{\rm FFE}=0=\nabla_{a}T_{\rm NFE}^{ab}$.  Then, the divergence of the bulk stress tensor is 
\begin{equation}
    \label{eq: bulk term divergence}
    \left(\nabla_aT_{\rm bulk}^{ab}\right)
    =
    m_a[T^{ab}]\delta(\Phi) .
\end{equation}
Next, consider $\nabla_{a}(s^{ab}\delta(\Phi))$. To treat this distributionally, introduce a test vector field $X^{a}$ with compact support containing the interface. 
Then, 
\begin{align}
    \int \nabla_{a}(s^{ab}\delta(\Phi))X_{b} \,\bm{\epsilon}
    &=-\int s^{ab}\delta(\Phi) \nabla_{a}X_{b}\,\bm{\epsilon}\notag
    \\
    &=-\int_{I} S^{AB}e_A{}^a e_B{}^b\nabla_a X_b\bm{\epsilon}_{I},
\end{align}
where $\bm{\epsilon}$ is a spacetime volume form and $\bm{\epsilon}_{I}$  is the volume form intrinsic to the interface manifold.  
On the interface, decompose
\begin{align}
    X_b=X_B e^B{}_b+X_\perp m_b,
\end{align}
where the coefficients of $e\indices{^B_b}$ and $m^{b}$ are given by
\begin{align}
    X_B=e_B{}^b X_b,\qquad
    X_\perp=m^bX_b. 
\end{align}
With the convention for \(K_{AB}\) above, this leads to 
\begin{equation}
    e_A{}^a e_B{}^b\nabla_a X_b
    =
    D_A X_B+K_{AB}X_\perp ,
\end{equation}
where \(D_A\) is the covariant derivative compatible with \(h_{AB}\), whose defining condition is $e_A{}^a e_B{}^b\nabla_a X_B=D_A X_B$ \cite{Wald:1984rg}.   After
integrating the first term by parts on \(I\), and assuming no boundary
contribution, we obtain
\begin{align}
     & \int \nabla_{a}(s^{ab}\delta(\Phi))X_{b} \,\bm{\epsilon}\notag
     \\
    &=
    \int_I \left[(D_A S^{AB})X_B-K_{AB}S^{AB}X_\perp\right]\,\bm{\epsilon}_{I}\notag
    \\
    &=
    \int \left[(D_A S^{AB})e_B{}^b-K_{AB}S^{AB}m^{b}\right]X_{b}\delta(\Phi)\,\bm{\epsilon}
\end{align}
Therefore, as a spacetime distribution,
\begin{equation}
    \label{eq: singular part divergence}
    \nabla_a(s^{ab}\delta(\Phi))
    =
    \left(D_A S^{AB}e_B{}^b-K_{AB}S^{AB}m^b\right)\delta(\Phi) .
\end{equation}
Using Eqs.~\eqref{eq: bulk term divergence} and \eqref{eq: singular part divergence}, the conservation of total stress-energy \(\nabla_aT^{ab}=0\) requires
\begin{equation}
    m_a[T^{ab}]
    +D_A S^{AB}e_B{}^b
    -K_{AB}S^{AB}m^b
    =0 .                                      \label{eq:surface-balance-vector}
\end{equation}
Projecting Eq.~\eqref{eq:surface-balance-vector} tangent to the interface using $e^B{}_b$
gives
\begin{equation}
    D_A S^{AB}
    =
    -e^B{}_b m_a[T^{ab}],
\end{equation}
which is Eq.~\eqref{eq: surface stress S eq1}.  Projecting the equation normally using $m_{b}$ gives
\begin{equation}
    K_{AB}S^{AB}
    =
    m_a m_b[T^{ab}],
\end{equation}
which is Eq.~\eqref{eq: surface stress S eq2}.  
If there is no surface stress, $S^{AB}=0$, then $m_{a}[T^{ab}]=0$ is recovered. Our discussion in Sec.~\ref{subsec:junction condition}, however, shows that the FFE/NFE interface is genuinely a stress-energy-carrying layer.

\section{Vacuum collision of triangular waves}\label{app: vacuum collision}
In the collision of two triangular waves given by Eq.~\eqref{eq:triangle-f}, a $P=0$ region forms. After that, the FFE description must be replaced by NFE. To determine where the $P=0$ region first appears, we consider an auxiliary vacuum problem in which the same wave packets are evolved by the vacuum Maxwell equations (\ref{eq: 1+1D Maxwell 1}) and (\ref{eq: 1+1D Maxwell 2}), i.e., the fields are simply given by the superposition of the incoming and outgoing waves. The vacuum and FFE+NFE evolutions agree up to the first appearance of the $P=0$ region.  However, it is not the physical continuation of FFE after the loss of magnetic dominance. Thus, the vacuum evolution should be thought of as a useful way to give a clear identification of the birth front that supplies the NFE region. 

Using reflection symmetry, we focus only on the $z\geq0$ region. 
We assume $A>1/2$ so that a region with $P\leq0$ can develop. 
The first phase of the vacuum collision is the region \(0<t<W/4\). During this time, the incoming and outgoing wave packets overlap only in the region $0\leq z\leq t$, and in the overlap region, both packets are in the rising branch of Eq.~(\ref{eq:triangle-f}). Using Eq.~(\ref{eq: P-FFE}), the invariant $P$ in this phase is given by 
\begin{equation}
    P=
    \begin{cases}
        1-\dfrac{t^2-z^2}{\tauo^2},
        &0\le z\le t,\\[8pt]
        1,
        &t\le z.
    \end{cases}                              \label{eq:P-early-piecewise}
\end{equation}
The next phase of the collision corresponds to \(W/4\le t\le W/2\). During this phase, the rising branch of the outgoing wave overlaps with the rising branch of the incoming wave for $0\leq z\leq W/2-t$ (rising-rising branch), and with the falling branch of the incoming wave for $W/2-t\leq z\leq t$ (rising-falling branch). Therefore, $P$ is given by
\begingroup
\begin{equation}
    P=
    \left\{
    \begin{array}{@{}ll@{\quad}l@{}}
        1-\dfrac{t^2-z^2}{\tauo^2},
        &0\le z\le W/2-t,
        \\
        1-\dfrac{(W-t-z)(t-z)}{\tauo^2},
        &W/2-t\le z\le t,
        \\
        1,
        &t\le z,
    \end{array}
    \right.                                      \label{eq:P-transition-piecewise}
\end{equation}
\endgroup
Now, consider where the $P=0$ surface is formed. By setting the first expression in Eq.~(\ref{eq:P-early-piecewise}) to zero, we find that $P=0$ is given by 
\begin{equation}
    z^2=t^2-\tauo^2.                       \label{eq:early-P-zero}
\end{equation}
For the $z>0$ region, this is simply
\begin{equation}
    \label{eq: z* P=0 surface}
    z^*(t)=\sqrt{t^2-\tauo^2} .
\end{equation}
This surface indeed lies in the overlap region, since \(z^*(t)<t\), and therefore gives a valid $P=0$ surface. 
Furthermore, this surface is spacelike since 
\begin{equation}
    \frac{\dd z^*}{\dd t}=\frac{t}{\sqrt{t^2-\tauo^2}}>1.
\end{equation}
The expression (\ref{eq: z* P=0 surface}) continues to be a valid $P=0$ surface as long as it remains in the rising-rising branch. Once it reaches the boundary between the rising-rising and rising-falling branches, the $P=0$ surface must be given by a different expression. Since the boundary is given by $z=W/2-t$, the transition time $t_{1}$ is determined by the equation
\begin{equation}
    z^*(t_1)=\frac{W}{2}-t_1,
\end{equation}
which gives
\begin{equation}
    t_1=\frac{W}{4}+\frac{\tauo^2}{W}.       \label{eq:t1}
\end{equation}
For \(t>t_1\), the $P=0$ surface lies in the rising-falling branch. To find the $P=0$ surface for $t>t_{1}$, one needs to set the middle expression of Eq.~(\ref{eq:P-transition-piecewise}) to zero, giving
\begin{equation}
    0=1-\frac{(W-t-z^*)(t-z^*)}{\tauo^2}.
\end{equation}
Solving this quadratic equation gives
\begin{equation}
\begin{aligned}
    z^*(t)
    &=\frac{W\pm\sqrt{(W-2t)^2+4\tauo^2}}{2}.
\end{aligned}
\end{equation}
The plus-sign root lies outside the overlap region; therefore, the $P=0$ surface for $t>t_{1}$ is
\begin{equation}
    z^*(t)=\frac{W-\sqrt{(W-2t)^2+4\tauo^2}}{2}.      \label{eq:middle-P-zero}
\end{equation}
This $P=0$ surface is timelike since
\begin{equation}
    \frac{\dd z^*}{\dd t}=\frac{W-2t}{\sqrt{(W-2t)^2+4\tauo^2}},
\end{equation}
whose absolute value never exceeds $1$. 
At $t=W/2$, the two waves overlap symmetrically in space. Thus, the later evolution is obtained by time reflection about $t=W/2$. The timelike $P=0$ surface transitions to the spacelike surface at a later time. This second transition time $t_{2}$ is given by
\begin{equation}
    t_2=W-t_{1}=\frac{3W}{4}-\frac{\tauo^2}{W}.
\end{equation}
Similarly, the final moment of $P=0$ is given by $W-\tau_{0}$, leading to the complete expression for the $P=0$ surface in the $z>0$ region: 

\begin{equation}
    z^*(t)=
    \begin{cases}
        \sqrt{t^2-\tauo^2},
        & \tauo\le t\le t_1,\\[6pt]
        \dfrac{W-\sqrt{(W-2t)^2+4\tauo^2}}{2},
        & t_1\le t\le t_2,\\[10pt]
        \sqrt{(W-t)^2-\tauo^2},
        & t_2\le t\le W-\tauo .
    \end{cases}                              \label{eq:P-zero-lens}
\end{equation}
The $z<0$ side is simply the reflection of this result. 
We can denote this spacelike birth front as $\Birth$ and parametrize it as
\begin{align}
    \Birth=
    \left\{
    z=\sqrt{s^{2}-\tau_{0}^{2}},\quad
    t=s
    \;\middle|\;
    s\in [\tau_{0},t_{1}]
    \right\}.                                         \label{eq:boundary-B appendix}
\end{align}
This surface serves as the initial data surface for the NFE region.

\bibliography{extendedFFE}
\bibliographystyle{utphys}

\end{document}